\documentclass[longauth]{aa}  
\usepackage[dvips]{graphicx}
\usepackage{txfonts}
\begin{document}

   \title{The zCOSMOS redshift survey}

   \subtitle{Influence of luminosity, mass and environment on the galaxy merger rate.
\thanks{Based on observations obtained at the European Southern Observatory (ESO) Very Large Telescope
(VLT), Paranal, Chile, as part of the Large Program 175.A-0839 (the zCOSMOS Spectroscopic Redshift
Survey)}
}

  \author{
 L. de Ravel  \inst{1,2} 
\and P. Kampczyk  \inst{3}
 \and O. Le F\`evre  \inst{1}
 \and S. J. Lilly  \inst{3}
  \and L. Tasca  \inst{4}
 \and L. Tresse  \inst{1}
  \and C. Lopez-Sanjuan  \inst{1}
 \and M. Bolzonella \inst{5}
 \and K. Kovac \inst{3}
 \and U. Abbas \inst{19}
 \and S. Bardelli \inst{5}
 \and A. Bongiorno \inst{6}
 \and K. Caputi \inst{2}
 \and T. Contini \inst{7}
 \and G. Coppa \inst{11,6}
 \and O. Cucciati \inst{1}
 \and S. de la Torre \inst{2}
 \and J.S. Dunlop \inst{2}
 \and P. Franzetti \inst{4}
 \and B. Garilli \inst{4,1}
 \and A. Iovino \inst{8}
 \and J.-P. Kneib \inst{1}
 \and A. M. Koekemoer \inst{18}
 \and C. Knobel \inst{3}
 \and F. Lamareille \inst{7}
 \and J.-F. Le Borgne \inst{7}
 \and V. Le Brun \inst{1}
 \and A. Leauthaud \inst{16}
 \and C. Maier \inst{3}
 \and V. Mainieri \inst{9}
 \and M. Mignoli \inst{5}
 \and R. Pello \inst{7}
 \and Y. Peng \inst{3}
 \and E. Perez Montero \inst{7}
 \and E. Ricciardelli \inst{10}
 \and M. Scodeggio \inst{4}
 \and J. D. Silverman \inst{3}
 \and M. Tanaka \inst{10}
 \and D. Vergani \inst{5}
 \and G. Zamorani \inst{5}
 \and E. Zucca \inst{5}
 \and D. Bottini \inst{4}
 \and A. Cappi \inst{5}
 \and C. M. Carollo \inst{3}
 \and P. Cassata \inst{17}
 \and A. Cimatti \inst{11}
 \and M. Fumana \inst{4}
 \and L. Guzzo \inst{8}
 \and D. Maccagni \inst{4,1}
 \and C. Marinoni \inst{12}
 \and H. J. McCracken \inst{13}
 \and P. Memeo \inst{4}
 \and B. Meneux \inst{6,15}
 \and P. Oesch \inst{3}
 \and C. Porciani \inst{3}
 \and L. Pozzetti \inst{5}
 \and A. Renzini \inst{10}
 \and R. Scaramella \inst{14}
 \and C. Scarlata \inst{3}
 }
%
%
%
%
  \institute{
 Laboratoire d'Astrophysique de Marseille, , UMR 6110 CNRS-Universit\'e de Provence, BP8, F-13376 Marseille Cedex 12, France
 \and
   Institute for Astronomy, University of Edinburgh, Royal Observatory, Edinburgh, EH93HJ, UK
 \and
 Institute of Astronomy, ETH Zurich, CH-8093, Zurich, Switzerland
 \and
 INAF - IASF Milano, Milan, Italy
 \and
 INAF Osservatorio Astronomico di Bologna, via Ranzani 1, I-40127, Bologna, Italy
 \and
 Max-Planck-Institut f¨ur Extraterrestrische Physik, D-84571 Garching b. Muenchen, Germany
 \and
 Laboratoire d'Astrophysique de Toulouse-Tarbes, Universite de Toulouse, CNRS, 14 avenue Edouard Belin, F-31400 Toulouse, France
 \and
 INAF Osservatorio Astronomico di Brera, Milan, Italy
 \and
 European Southern Observatory, Karl-Schwarzschild-Strasse 2, Garching, D-85748, Germany
 \and
 Dipartimento di Astronomia, Universita di Padova, Padova, Italy
 \and
 Dipartimento di Astronomia, Universit´a di Bologna, via Ranzani 1, I-40127, Bologna, Italy
 \and
 Centre de Physique Theorique, Marseille, Marseille, France
 \and
 Institut d'Astrophysique de Paris, UMR 7095 CNRS, Universit´e Pierre et Marie Curie, 98 bis Boulevard Arago, F-75014 Paris, France
 \and
 INAF, Osservatorio di Roma, Monteporzio Catone (RM), Italy
 \and
 Universit¨ats-Sternwarte, Scheinerstrasse 1, D-81679 Muenchen, Germany
\and
Berkeley Lab \& Berkeley Center for Cosmological Physics, University of california, Lawrence Berkeley National Lab., 1 cyclotron road, MS 50-5005,Berkeley, CA
\and 
Dept. of Astronomy, University of Massachusetts at Amherst
\and
Space Telescope Science Institute, 3700 San Martin Dr, US Baltimore MD 21218-2410, United States
\and
INAF - 0sservatorio Astronomico di Torino, via Osservatorio 20, I-10025 Pino Torinese (TO), Italy
 \\
 \email{ldr@roe.ac.uk}
 }

   \date{Received ; accepted }

 
  \abstract
   {The contribution of major mergers to galaxy mass assembly along
    cosmic time is an important ingredient to the galaxy evolution
    scenario.}
   {We aim to measure the evolution of the merger rate for both luminosity/mass selected galaxy samples and investigate its dependence with the local environment.}
   {We use a sample of 10644 spectroscopically observed galaxies from the zCOSMOS redshift survey to identify pairs of galaxies destined to merge, using only pairs for which the velocity difference and projected separation of both components with a confirmed spectroscopic redshift indicate a high probability of merging.}
   {We have identified 263 spectroscopically confirmed pairs with $r_p^{max} = 100 h^{-1}\ kpc$. We find that the density of mergers depends on luminosity/mass, being higher for fainter/less massive galaxies, while the number of mergers a galaxy will experience does not depends significantly on its intrinsic luminosity but rather on its stellar mass. We find that the pair fraction and merger rate increase with local galaxy density, a property observed up to redshift $z\sim1$.}
   {We find that the dependence of the merger rate on the luminosity or 
      mass of galaxies is already present up to redshifts $z \sim 1$, and that the evolution of 
       the \textit{volumetric} merger rate of bright (massive) galaxies is
    relatively flat with redshift with a mean value of $\sim 3 \times 10^{-4} (8 \times 10^{-5}\ respectively)\ mergers\ h^3 Mpc^{-3} Gyr^{-1}$. The dependence of the
    merger rate with environment indicates that dense 
    environments favors major merger events as can
    be expected from the hierarchical scenario. The environment
    therefore has a direct impact in shapping-up the mass function and 
    its evolution and therefore plays an important role
    on the mass growth of galaxies along cosmic time.}

\maketitle
%

\section{Introduction}

In the last years, with large observational programs reaching earlier stages in the life of the Universe, the importance of mergers in the evolution of galaxies has become an observational fact. This is not a surprise as, in the current hierarchical structure formation paradigm, the process of coalescence of dark matter halos should strongly impact the mass assembly in galaxies living within these halos.

In the hierarchical picture, the halo merging tree history can be quantified by a halo merger rate, measuring the growth of mass per average mass in a representative volume of the Universe, and its evolution with redshift $\propto (1+z)^{3/2}$ (\cite{kitz}), but observational constraints come from the observations of galaxy-galaxy mergers which may not directly follow the DM merger rate evolution. In the last decade, a number of studies have attempted to measure the galaxy merger rate and its evolution using a variety of techniques, producing a wide range of  apparently inconsistent values (e.g. \cite{P97}; \cite{P00};  \cite{lefevre00};\cite{P02};\cite{conselice03}; \cite{bundy05}; \cite{kartaltepe07}; \cite{kampczyk}; \cite{lotz08}; \cite{bundy09}), but difficult to compare because of different selection functions and methodologies. It is only with recent large spectroscopic redshift surveys that robust estimate of the merger rate have become possible using a systematic confirmation of physically merging pairs, followed by a physical analysis of their properties. In the DEEP2 survey, Lin et al. (2008) observed that the evolution of the merger rate has been flat with redshift. In the Vimos VLT Deep Survey (VVDS) sample extending to fainter luminosities, de Ravel et al. (2009) have demonstrated that the merger rate evolution strongly depends on the luminosity or mass of the galaxies, with brighter galaxies showing a flatter merger rate evolution consistent with what is observed in the DEEP2 sample (up to $z=1.4$ at least), while fainter galaxies with $M_B \leq -18$ show a stronger evolution parametrized by $(1+z)^{2.20 \pm 0.77}$. In itself this result reconciles some of the apparently discrepant measurements obtained previously, which can now be understood in the framework of galaxy samples with different luminosity or mass (see also Lopez-Sanjuan et al. (2010a)). 

In parallel, the theoretical modeling of structure formation and large numerical simulations are produce key advances in our understanding of the role of galaxy-galaxy merging and associated timescales in a cosmological context (e.g. \cite{lotz08}). The merging timescale is an important parameter when deriving the merging rate from pair count observations, and it has been classically assumed to be roughly half a billion years for massive galaxies with roughly equal mass (e.g. \cite{P00};\cite{P02};\cite{lin04}). In a detailed analysis of the Millenium simulation Kitzbichler \& White (2008) showed that the actual merger timescale for a pair of galaxies may be considerably larger for all masses than this previously assumed value, reaching an excess of one billion years. Therefore, merging timescale probably remains the largest input of uncertainty in these kind of calculations.

However, these simulations enable to investigate more accurately the history of galaxies with respect to their stellar mass accretion. There are increasingly clear evidences that the galaxy mass assembly process is in a very active phase between $1\leq z \leq 2$ as observed from the evolution of the stellar mass density (e.g. \cite{bundy04}; \cite{arnouts}; \cite{ilbert10}), and one would naturally expect merging to be a major if not dominant contributor. However, there is currently a debate as to the importance of merging in the build up of galaxies, and in particular on how mergers impact the formation of early-type massive galaxies at high redshift. Studies of the velocity fields with 3D integral field spectroscopy of bright $z \sim 2$ galaxies find a number of self supported rotating discs or high velocity dispersions galaxies which may be assembled from gas accretion via cold flows along streams of the cosmic web fueling the star formation rather than from merging processes (\cite{genzel09}; \cite{dekel09}). However, major merging events represent about one third of galaxies observed in 3D spectroscopy of galaxies with z $\sim 1-3$ (\cite{epinat}; \cite{forster}; \cite{law}), indicating that merging is indeed very active at these epochs. 

It is clear now that mergers produce strong modifications of galaxies properties. The most obvious effect is a change in the morphology. Indeed, major mergers have been proposed to be the main mechanism that transform disk-like galaxies into spheroidals. The intrinsic nature of galaxies themselves could also be affected by a merging process : recent studies have shown direct evidence of merger-induced Active Galaxy Nuclei activity (\cite{ramos}; \cite{silverman}). The color of galaxies is also strongly affected by major merging processes. During the merger itself, inter-stellar clouds may collapse and many new stars are formed in a starbust process producing enhanced star formation rates (\cite{hopkins06}; \cite{bastian}; \cite{deravel}; \cite{darg09}; \cite{Bridge10}). This enhancement of star formation can depletes the galaxy gas reservoir on a short timescale leaving the remnant  "red" and "dead". We know that a color bimodality with a red sequence is also already present at $z \sim 1-1.5$ and that blue galaxies show only weak evolution in number density compared to the number of red galaxies that has doubled/tripled since $z \sim 1$ (e.g. \cite{ilbert06}; \cite{arnouts}). 

It therefore becomes increasingly clear that there is a chain of events that lower the number density of intermediate late type galaxies to build up more early type galaxies. The best way to probe the formation of galaxies that populates the red sequence is to identify their progenitors. The so-called \textit{k+a} galaxies have been suggested to be one of them. These galaxies show strong Balmer absorption lines, indicative of an intense star formation epoch within the last few billion years but no-ongoing star formation. Multiple origins of \textit{k+a} have been proposed (e.g. \cite{vergani}) but these caracteristics are those one could expect from a galaxy that experienced a recent \textit{wet} merger (\cite{wild}). Knowing the importance of \textit{wet} mergers in the formation of \textit{k+a} galaxies could therefore be a link towards the understanding of the formation of early-type galaxies.
However, mergers also happen between galaxies within the red sequence. These are called \textit{dry} mergers (\cite{khochfar03}; \cite{bell06b}; \cite{dimatteo07}; \cite{scarlatta07}; \cite{lin08}; \cite{lin10}; \cite{robaina}). If one wants to understand the role of mergers in galaxy evolution, it is therefore crucial to understand when and where these different kind of mergers happen. Recent studies have shown that the merging activity is more active at high redshift (\cite{depropris10}; \cite{eliche10b}). The bulk of mass assembly through mergers is thought to lie between $1\leq z\leq 2$. 

Galaxy merger rate depends on stellar mass (e.g. \cite{conselice06}; \cite{deravel}; ) and since massive galaxies are more likely to live in dense regions, merger rate should depend on environment as well. By definition galaxy mergers are expected to take place in relatively high density regions. It has been shown, locally that , indeed, mergers density distribution peaks in average these intermediate/high density regions (e.g. \cite{mcintosh08}; \cite{darg09}; \cite{perez}; \cite{ellison10}) corresponding roughly to galaxy groups where the velocity dispersion is not as strong as in galaxy clusters. In a recent paper, Lin et al. (2010) have shown that this trend seems to remains at higher redshift ($0.75 \leq z \leq 1.2$). They shown a clear dependence of the merging rate with the local density and found that most of this dependence comes from the dry mergers due to the increased population of red galaxies in these dense environments. Our paper aims to give a general picture of the role of density in the merger rate evolution between $z=1$ down to $z=0.2$.

Therefore, we report here on the use of the zCOSMOS spectroscopic redshift survey of bright galaxies up to $z \sim 1$ to establish a new measurement of the merger rate evolution from a count of real physical pairs and to test the relationship between the merger rate and the local environment. The layout of the paper is as follows. We present the zCOSMOS sample in Section~\ref{sample}. The methodology to identify close pairs with both members of the pairs having a confirmed
spectroscopic redshift, and establish the pair sample, is presented in Section~\ref{pairs}. We detail the corrections applied in Section~\ref{selection_effects} and present the redshift evolution of the pair fraction in Section~\ref{pair-fraction}. The estimate of merger time-scales and the evolution of the merger rate based on the pair fraction is presented in Section~\ref{merger-rate}. We study the influence of the luminosity on the merger rate in Section~\ref{luminosity}. Then, in Section~\ref{environment}, we study the impact of environment on the galaxy merger rate. Eventually, we discuss and conclude in Section~\ref{conclusion}.

Throughout this paper we assume a flat $\Lambda CDM$ cosmology with $\Omega_m = 0.25$, $\Omega_{\Lambda}=0.75$ and $H_0 = 70\ km s^{-1}Mpc^{-1}$. All magnitudes are given in the
AB system.

   \section{zCOSMOS Bright sample}
\label{sample}
 \subsection{Spectroscopy}
\label{spectro}
zCOSMOS (\cite{Lilly}) is a large on-going spectroscopic redshift survey within the COSMOS (\cite{scoville}) field performed with the VIMOS spectrograph (\cite{lefevre03}) on the European Southern  Observatory's Very Large Telescope (ESO-VLT). In this analysis we use the bright part of this survey, called the zCOSMOS-bright \textit{10k},  which covers the COSMOS HST/ACS field (\cite{koekemoer07}). The strategy of multiple passes of the VIMOS spectrograph will allow to reach a high galaxy sampling rate of ~70\% when complete. The \textit{10k} sample has a mean sampling rate of ~33\%. Spectra of galaxies have been obtained using the Red Medium-Resolution grism (MR) with a spectral resolution R=600 covering the wavelength range 5550 to 9650 \AA{}. This is a pure magnitude selected sample with $I_{AB} \leq 22.5$. A total of 10\,644 galaxies have been observed with VIMOS in multi-slit mode, and the data have been processed using the VIPGI data proccesing pipeline (\cite{scodeggio}). A spectroscopic flag has been assigned to each galaxy providing an estimate of the robustness of the redshift measurement (\cite{Lilly}). If a redshift has been measured, the corresponding spectroscopic flag value can be 1, 2, 3, 4 or 9. Flag $1$ meaning that the redshift is $\sim 50\%$ secure and flag 4 that the redshift is $\sim 99\%$ secure. Flag 9 means that the redshift measurement rely on one single narrow emission line ($OII$ or $H\alpha$ mainly). An information about the consistency between photometric and spectroscopic redshift has also been included as a decimal in the spectroscopic flag. In this study we select the highest reliable redshifts, i.e. with confidence class 4.5, 4.4, 3.5, 3.4, 9.5, 9.3, 2.5 for main targets and for serendipitous galaxies within the same magnitude limit and corresponding confidence classes. This provides a sample of $\simeq6800$ galaxies in the redshift range $0.2 \leq z \leq 1$ with a confirmation rate of 99.5\%, with a peak in the redshift distribution at $z \sim 0.6$ and a velocity accuracy estimated to be $ \sim 100\ km s^{-1}$. 
Spectroscopic redshifts and photometry (see Section~\ref{photo}) have been used to compute absolute magnitude using the code \textit{ALF} (see Ilbert et al. (2005) for further details).

\subsection{Photometry}
\label{photo}
The input target catalogue for zCOSMOS-bright is based primarily on the F814W magnitudes derived from the HST-ACS images (\cite{koekemoer07}). For the small areas where the HST data is absent or compromised by diffraction spikes etc, these are supplemented by photometry from a high spatial resolution i-band CFHT image which has diffraction spikes at different position angles (See Lilly et al. (2007) for further details). The statistical reliability of the spectroscopic redshifts in the various Confidence Classes is assessed by the comparison of repeat observations of a sample of 600 galaxies and by the comparison to photometric redshifts computed from codes taking into account stars, galaxies or AGN (\cite{ilbert08}; \cite{salvato09}; \cite{feldmann08}). The redshift distribution $N(z)$ of both photometric and spectroscopic samples highlight two structures at $z \sim 0.35$ and $z \sim 0.7$. Moreover, in order to select the same regions probed with the spectroscopic sample, we have extracted from the original parent catalogue all the galaxies included in the combined footprint of the VIMOS quadrants.


   \section{Spectroscopic pairs in the zCOSMOS\textit{-bright} sample}
\label{pairs}

\begin{figure}[htbp!]
 \centering
 \input{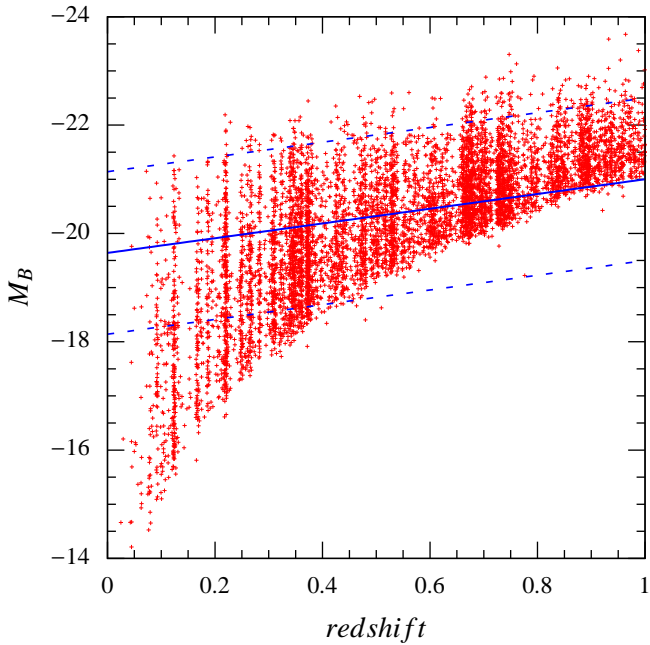}
 \caption{The absolute magnitude in $B$ band is plotted versus redshift for all galaxies of the \textit{10k}-sample. The solid blue line is the absolute magnitude cut we applied to compare galaxies of similar luminosities throughout the redshift range of our survey. We assumed a passive evolution with $M_B < -19.64 -Q(z)$. Dotted lines show $\pm 1.5$ magnitudes around this cut.}
 \label{mag-distrib}
 \end{figure}

We identify close galaxy pairs from our spectroscopic sample by measuring three key quantities for each galaxy couple: the projected physical separation $r_p$, the velocity difference along the line-of-sight $\Delta v$, and the absolute magnitude difference in the B-band $\Delta M_B$. Thus, close pairs are defined to have $r_p \leq r_p^{max}$ and $\Delta v \leq \Delta v^{max}$. Thanks to spectroscopically confirmed redshifts, both projected and velocity separations between two galaxies are accurate enough for our study. The last criterion $\Delta M_B^{max}$, the maximal absolute magnitude difference, enable us to focus on major mergers. These particular type of mergers are those involving galaxies of comparable luminosities or stellar masses. 
As we aim to compare the merger rate of galaxies of comparable luminosities throughout the redshift range of our survey, we use an absolute magnitude cut of $M_B=-21$ at $z = 1$. This ensures that we study a complete sample at this redshift, and we apply a passive luminosity evolution following the evolution of $M_B^*$ as measured from the luminosity function of our sample (\cite{zucca09}). This translates into a luminosity  threshold evolving as $M_B^{lim}(z) = M_B(z=0) - Q \times z$ with $Q=1.36$ (see Fig~\ref{mag-distrib}).

Two different approaches, leading to different definitions of the pair fraction, have been followed. 
First, one can define a pair as a system where both members are brighter than $M_B^{lim}$. This approach reduces greatly the statistics because the dynamic range decreases with redshift. However, this prevents from applying any luminosity dependent correction since both galaxies lies in our volume limited sample. This approach can there be seen as a lower limit to the pair fraction, specially at $z \sim 1$ where the reduced dynamic range reduces the observable pair statistics.

The other way to define a pair is to apply the absolute magnitude threshold only for one galaxy, and search for companions satisfying the magnitude difference $\Delta M_B$, meaning that some of the companions will have an absolute magnitude fainter than the luminosity threshold. Therefore, this technique requires to correct for companions we can't see in our volume limited sample at the highest redshift probed here. This correction relies on the luminosity function and increases with redshift since one reaches the limit of our volume-limited sample (see section 4.3 for more details).

These two approaches answer different questions. If one define pairs to have both their members above the luminosity threshold, the pair fraction will tell us how many galaxies with both $M_B \leq M_B^{lim}$ are in interaction. If one define pairs to have only one galaxy with $M_B \leq M_B^{lim}$, the pair fraction will reflects the number of galaxies with $M_B \leq M_B^{lim}$ that have a companion within the 1.5 magnitude range defined for the study of major mergers  (in the $B$ band). In the following we will present results using the second technique (where only one member of the pair has to be brighter than the luminosity threshold), however, a detailed description using the first technique will be presented in Kampczyk et al. (in prep).

Looking for major mergers, i.e. equivalent luminosity/mass pairs with a high merging probability, imposes to use a criteria based on luminosity or mass difference between the two galaxies. In the case of our luminosity selected sample, we use a criterium based on the absolute magnitude difference in the B-band  : $\Delta M_B^{max} = 1.5$. This translates, for mass selected samples, in a stellar mass difference less than 1:4. Futhermore, we impose to our pairs to have a velocity difference along the line of sight of $\Delta v \leq\ 500\ km/s$. Table~\ref{table_Npaires} gives the number of pairs in each redshift bin considered for maximal projected separations of $r_p^{max}=20, 30,\ 50$ and $100\ h^{-1} kpc$. In the following, we will ignore the redshift bin $0 \leq z \leq 0.2$ because of the small volume sampled resulting in a small number of pairs even for high separations. In total, 39 pairs were found across the entire field for $r_p^{max} = 20\ h^{-1} kpc$ and $0.2 \leq z \leq 1.0$. We list the properties of these identified close pairs with $r_p^{max}=20\ h^{-1} kpc$ in Table~\ref{list}. This number increases to 263 pairs when we allow the  separation to reach $r_p^{max} = 100\ h^{-1} kpc$.

\begin{table}[ht]
\center
\caption{Number of pairs with $\Delta M_B^{max} = 1.5$ and $M_B < -19.64 -Q(z)$.}
\label{table_Npaires}
\centering
\begin{tabular}{c c c c c}

  & $20h^{-1}\ kpc$ & $30h^{-1}\ kpc$ & $50h^{-1}\ kpc$ & $100h^{-1}\ kpc$\\ \hline\hline \
\\
$0.2 \leq z \leq 0.4$ & 6 & 10 & 21 & 49\\ 
$0.4 \leq z \leq 0.6$ & 5 & 14 & 22 & 57\\ 
$0.6 \leq z \leq 0.8$ & 19 & 25 & 45 & 110\\
$0.8 \leq z \leq 1.0$ & 9 & 17 & 21 & 47\\
total & 39 & 66 & 109 & 263\\ \hline\hline
\end{tabular} 
\end{table}

 \begin{table*}[ht!]
 \center
 \caption{List of the 39 spectroscopic pairs with $r_p^{max}= 20h^{-1} kpc$, $\ \Delta v^{max} = 500\ km/s\ \ $\ and\ \  $\Delta M_B^{max} = 1.5$ mag selected in the bright $M_B(z=0) \leq -19.64$ sample. The total number of pairs increases to 263 when $r_p^{max}= 100h^{-1} kpc$.}
 \label{list}
 \centering
 \begin{tabular}{c|c c c c c c c}
 \hline \hline \\
 Pair number & Id1 & Id2 & $z_{mean}$ & $r_p\ ( h^{-1} kpc )$ & $\Delta v\ (km/s)$ & $\Delta M_B$ & $\theta\ (")$\\ \hline\hline \\
 1 & 804056 & 804061 & 0.9033 &  16.4 &  94.6 &  0.50 & 2.9\\ 
 2 & 806888 & 806891 & 0.5064 &  17.7 &  19.9 &  1.31 & 4.0\\ \
 3 & 811012 & 811015 & 0.8385 &  16.9 &  32.6 &  0.18 & 3.1\\ \
 4 & 811488 & 811491 & 0.6230 &   7.6 & 129.4 &  0.06 & 1.6\\ \
 5 & 811990 & 811991 & 0.7363 &   7.2 &  86.4 &  0.20 & 1.4\\ \
 6 & 812557 & 812561 & 0.5384 &  16.7 & 156.0 &  0.64 & 3.7\\ \
 7 & 813041 & 813043 & 0.8695 &  15.3 & 272.8 &  0.38 & 2.7\\ \
 8 & 813227 & 813229 & 0.3716 &  16.8 & 240.6 &  0.64 & 4.6\\ \
 9 & 813462 & 813466 & 0.6936 &  19.8 & 354.3 &  0.80 & 3.9\\ \
 10 & 817147 & 817153 & 0.9642 &  17.9 & 152.7 &  1.19 & 3.1\\ \
 11 & 817797 & 817800 & 0.6714 &  19.7 & 161.5 &  0.60 & 3.9\\ \
 12 & 817830 & 817832 & 0.7263 &   6.9 & 156.4 &  0.30 & 1.3\\ \
 13 & 817865 & 817867 & 0.7230 &  14.0 &  69.6 &  1.13 & 2.7\\ \
 14 & 818361 & 818366 & 0.6224 &  10.2 & 203.4 &  0.49 & 2.1\\ \
 15 & 819899 & 819900 & 0.8919 &   9.1 & 111.0 &  0.67 & 1.6\\ \
 16 & 820084 & 820087 & 0.6464 &  18.6 &  72.9 &  0.58 & 3.8\\ \
 17 & 822902 & 822904 & 0.8366 &  17.1 & 277.7 &  0.89 & 3.1\\ \
 18 & 823078 & 823080 & 0.4062 &  19.3 & 149.3 &  0.88 & 5.0\\ \
 19 & 824472 & 824476 & 0.6688 &  19.0 & 125.8 &  1.45 & 3.8\\ \
 20 & 824655 & 824662 & 0.6603 &  13.0 & 253.0 &  0.24 & 2.6\\ \
 21 & 824745 & 824746 & 0.8479 &  12.1 & 146.1 &  0.53 & 2.2\\ \
 22 & 825330 & 825338 & 0.4728 &  19.3 &  81.5 &  0.85 & 4.6\\ \
 23 & 825904 & 825908 & 0.3448 &   7.1 & 223.1 &  0.61 & 2.1\\ \
 24 & 826452 & 826453 & 0.3542 &   5.5 & 221.5 &  0.69 & 1.6\\ \
 25 & 827363 & 827365 & 0.6765 &   6.1 &  35.8 &  0.11 & 1.2\\ \
 26 & 831183 & 831185 & 0.6055 &   3.8 & 112.1 &  0.91 & 0.8\\ \
 27 & 831248 & 831249 & 0.6842 &  18.2 & 285.0 &  0.04 & 3.6\\ \
 38 & 831776 & 831781 & 0.8369 &   9.7 & 228.6 &  0.07 & 1.8\\ \
 29 & 832137 & 832145 & 0.7280 &  16.2 & 277.8 &  0.77 & 3.1\\ \
 30 & 832428 & 832433 & 0.6587 &  19.4 &  90.4 &  1.22 & 3.9\\ \
 31 & 835862 & 835863 & 0.4027 &   4.0 &  21.4 &  1.36 & 1.0\\ \
 32 & 837325 & 837327 & 0.2193 &  16.3 & 172.2 &  0.97 & 6.5\\ \
 33 & 837599 & 837602 & 0.3760 &   8.7 &  21.8 &  1.19 & 2.4\\ \
 34 & 840266 & 840268 & 0.9598 &   9.1 & 122.5 &  0.43 & 1.6\\ \
 35 & 844187 & 844188 & 0.6233 &  12.3 & 166.3 &  0.93 & 2.5\\ \
 36 & 844480 & 844484 & 0.6112 &   8.9 & 223.4 &  1.50 & 1.9\\ \
 37 & 844761 & 844762 & 0.7507 &  15.1 & 257.0 &  0.33 & 2.9\\ \
 38 & 845206 & 845207 & 0.3980 &   8.7 & 321.9 &  0.59 & 2.3\\ \
 39 & 847268 & 847274 & 0.7022 &  13.5 &  52.9 &  0.26 & 2.6\\ \hline\hline
 
 \end{tabular} 
 \end{table*}

\section{Accounting for selection effects}
\label{selection_effects}

The pair fraction is basically the number of pairs over the number of galaxies satisfying the same selection criteria. However, because of the selection function, estimates of pair fractions from spectroscopic surveys require a serie of corrections which are described in this section. 
\subsection{Angular completeness}
\label{angular-comp}
\begin{figure}[htbp!]
 \begin{center}
    \input{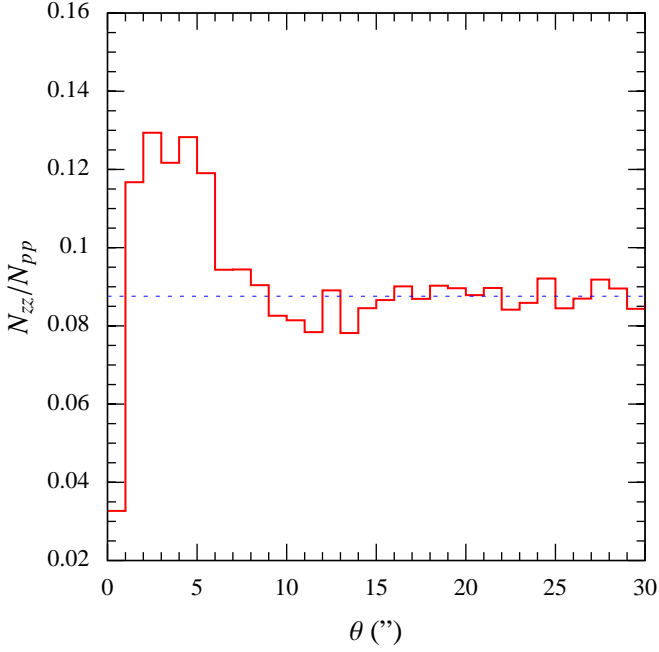}
   \caption{Completeness as a function of angular separation $\theta$. This completeness is defined as the number of spectroscopic pairs $N_{zz}$ over the number of photometric pairs $N_{pp}$. The dotted line represent the average value at large scales with $\theta \geq 15\ ''$. The bump for separations $1 \leq \theta \leq 10\ ''$ is due to objects falling into the slit by chance. }
      \label{angular-correction}
 \end{center}
  \end{figure}

We have first to investigate the angular completeness of the survey. With a multi slits instrument like VIMOS, it is not possible to observe two objects too close on the plane of the sky because of the slit overlap this would produce. Furthermore, as it is necessary to include in each slit the sky background on each side of a selected target to enable for accurate sky subtraction, a minimum separation of about 3 arcseconds between targets is imposed on a single mask observation. After several VIMOS observations of the same field using different slit masks close pairs are observed in a random fashion, partially erasing this minimum separation cut, and the fundamental limitation on pair separation is then set by the image quality of the images used to produce the photometric catalog, or about 1 arcsec. We therefore underselect galaxy pairs at small angular separations less than a few arcseconds. To estimate this bias, we compare the number of photometric pairs in the parent photometric catalogue with the number of pairs we find in the spectroscopic \textit{10k}-sample, and we use the ratio between the number of spectroscopic pairs $N_{zz}$ and the number of photometric pairs $N_{pp}$ as a function of the angular pair separation $\theta$ as shown in Figure\ref{angular-correction} to correct for this bias. The evident lack of spectroscopic pairs at small separations ($\leq 1\ ''$) corresponds to the image quality of the photometric images. For $1 \leq \theta \leq 10\ ''$, objects with the right magnitude can fall in the slit by chance, and this leads to an increase of this ratio. For $ \theta \geq 15\ ''$ the ratio is pretty constant with $ a = \frac{N_{zz}}{N_{pp}}(\theta \geq 15\ '') = 0.0876$. Therefore, we assign to each pair $k$ a weight $\omega_{\displaystyle \theta}^{\displaystyle k}$ which depends on the pair separation : $$\omega_{\displaystyle \theta}^{\displaystyle k} = \frac{\displaystyle a}{\displaystyle \theta_{k}}.$$

  \subsection{Spectroscopic completeness}
\label{spectro-comp}
The zCOSMOS\textit{-bright} survey is not spectroscopically complete since about one third of galaxies have been targetted by the VIMOS spectrograph and measuring accurate redshifts has not been possible for all galaxies. This incompleteness, translated into a Target Sampling Rate (TSR) and a Spectroscopic Success Rate (SSR), must be taken into account when computing pair statistics and we use the same method  as described in de Ravel et al. (2009)  to correct for them. To account for the TSR, we apply a weight $$\omega_{TSR}=\frac{\displaystyle N_{\displaystyle g}^{\displaystyle spectro}}{\displaystyle N_{\displaystyle g}^{\displaystyle photo}},$$ where $N_{\displaystyle g}^{\displaystyle spectro}$ is the total number of galaxies observed in the spectroscopic survey and $N_g^{photo}$ is the number of galaxies in the parent photometric catalogue (see Lilly et al (2007) for further details). To correct for the SSR, we apply a weight $$\omega_{SSR}=\frac{\displaystyle N_{\displaystyle g}^{\displaystyle spectro,flag}}{\displaystyle N_{\displaystyle g}^{\displaystyle spectro}},$$ where $N_{\displaystyle g}^{\displaystyle spectro,flag}$ is the number of galaxies in the spectroscopic survey  that have a reliable spectroscopic flag (see Section~\ref{spectro}). The combined completeness weight for each galaxy $i$ is then written as : $$\omega_{\displaystyle comp}^{\displaystyle i}=\omega_{TSR}^{\displaystyle -1} \times \omega_{SSR}^{\displaystyle -1}\ .$$ It depends on redshift and apparent magnitude (see \cite{zucca09} for further details). Furthermore, we assign each pair $k$ with $$\omega_{\displaystyle p,comp}^{\displaystyle k}=\omega_{\displaystyle comp}^{\displaystyle i} \times \omega_{\displaystyle comp}^{\displaystyle j}\ ,$$ where $\omega_{\displaystyle comp}^{\displaystyle i}$ and $\omega_{\displaystyle comp}^{\displaystyle j}$ are the spectroscopic completeness weights of the two galaxies in pair.

A few hundred X-ray identifications and a handful of other targets were included in the mask designs as compulsory targets (see \cite{Lilly} for further details). Since these have a much greater chance of being included in the spectroscopic masks we include a de-weighting scheme to correct for their higher chance of having been observed spectroscopically. The mean sampling rate of these objects being ~2 times that of the other random targets, we assigned manually an additionnal  specific weight of $0.5$ to all these compulsory objects.

\subsection{Luminosity incompleteness}
\label{luminosity-comp}
 Figure~\ref{mag-distrib} shows the distribution of absolute magnitudes in the B-band as a function of redshift. Here, we are looking for pairs for which at least one of the member is brighter than the absolute magnitude cut defined (see Section~\ref{pairs}) : $M_{B} \leq M_B(z=0) - Q \times z$. At the high redshift end of our survey, we are not able to identify pairs when the fainter companion has an apparent magnitude fainter than our $I_{AB}=22.5$ limit, therefore artificially lowering the number of pairs. To take this into account, we follow de Ravel et al. (2009) by computing for each galaxy $i$ a weight $\omega_{mag}^{i}$ using the luminosity function of this survey (\cite{zucca09}) to estimate the number of galaxies thus missed. For each galaxy $i$, we derive $M_{sup}^i = M_B^i + \Delta M_B^{max}$ that a companion is allowed to have to satisfy the search criteria. Using the survey limit $I_{AB}=22.5$, we derive for each galaxy $i$ at redshift $z_i$ the maximum absolute magnitude reached by the survey at that redshift $M_{sel}^i$. We then assign a weight for each galaxy $i$:
 
  $$\omega_{mag}^{i}(M_B,z) = \left\{\begin{array}{ll}
   1 & \mbox{if $M_{sup}^{i} \leq M_{sel}^{i}$}\\
  \\
   \frac{\displaystyle \int_{\displaystyle -\infty}^{\displaystyle M_{sup}^{i}} \Phi(M) dM}{\displaystyle \int_{\displaystyle -\infty}^{\displaystyle M_{sel}^{i}} \Phi(M) dM} & \mbox{if $M_{sup}^{i} > M_{sel}^{i}$\ ,}
   \end{array}\right.$$
 
 where $\Phi(M)$ is the global luminosity function.  \ \\
 The median weight $<\omega_{mag}>$ increases as a function of redshift : $<\omega_{mag}>\ =\ 1$, $1.21$, $1.58$ and $2.24$ within $0.2\leq z\leq 0.4$, $0.4\leq z\leq 0.6$, $0.6\leq z\leq 0.8$ and $0.8 \leq z \leq 1.0$ respectively. At this point it should be noticed that this correction is only based on the luminosity function with the caveat that this does not include any information about clustering properties of objects at small scales. However, Patton \& Attfield (2008, Fig 7 in their paper) have shown that, at least in the local Universe, at a given redshift, the co-moving number density of close companions does not depends strongly on absolute magnitude below over two orders of magnitude below $M_*$. Therefore, our correction based on the ratio between number densities of population with at most 1.5 magnitude difference should not introduce a significant bias. 

We therefore assign to each pair $k$ of two galaxies $i$ and $j$ the weight : $\omega_{\displaystyle p,mag}^{\displaystyle k} =\omega_{\displaystyle mag}^{\displaystyle i} \times \omega_{\displaystyle mag}^{\displaystyle j} $

\subsection{Combining weights}
\label{combining}
 The corrected number of pairs $N_p^{corr}$ and galaxies $N_g^{corr}$ is therefore computed as :
 
 $$N_p^{corr}(z) =  \sum_{\displaystyle k=1}^{\displaystyle N_{p,obs}} \omega_{\displaystyle p,comp}^{\displaystyle k} \times \omega_{\displaystyle p,mag}^{\displaystyle k} \times \omega_{\displaystyle \theta}^{\displaystyle k}$$
 and the corrected number of underlying galaxies is 
 $$N_g^{corr}(z)=  \sum_{\displaystyle i=1}^{\displaystyle N_{g,obs}} \omega_{\displaystyle comp}^{\displaystyle i} \times \omega_{\displaystyle mag}^{\displaystyle i}.$$
 
 The parent galaxies from which the pairs have been identified are those galaxies matching the absolute magnitude criterion within the overall \textit{10k}-catalog. Using these corrected numbers, we can estimate the pair fraction $f_p(z)$ in each redshift bin as :
 
 $$f_p(z)=\frac{\displaystyle N_{\displaystyle p}^{\displaystyle corr}(z)}{N_{\displaystyle g}^{\displaystyle corr}(z)}.$$

Assuming that merging systems are dominated by couples rather than higher multiplets, which is the case in our sample, the pair fraction as given here represents the fraction of pairs with respect to the underlying galaxies. Therefore, the probability that an individual galaxy with $M_B \leq M_B^{lim}$ has a close companion with $\Delta M_B \leq \Delta M_B^{max}$ is twice this fraction.

   \section{Pair fraction evolution...}
\label{pair-fraction}

\subsection{... in luminosity selected samples}

 The weighting scheme described in Section~\ref{combining} allows to estimate the evolution of the pair fraction $f_p(z)$ with redshift. We perform the calculations for different sets of projected separations ($r_p^{max}\ =\ 20$, $30$, $50$ and $100 h^{-1}\ kpc$) for galaxy pairs with $\Delta v^{max}=500\ km s^{-1}$ and containing at least one galaxy brighter than $M_B^{lim}$. 

\begin{figure*}[htbp!]
  \begin{center}
  \input{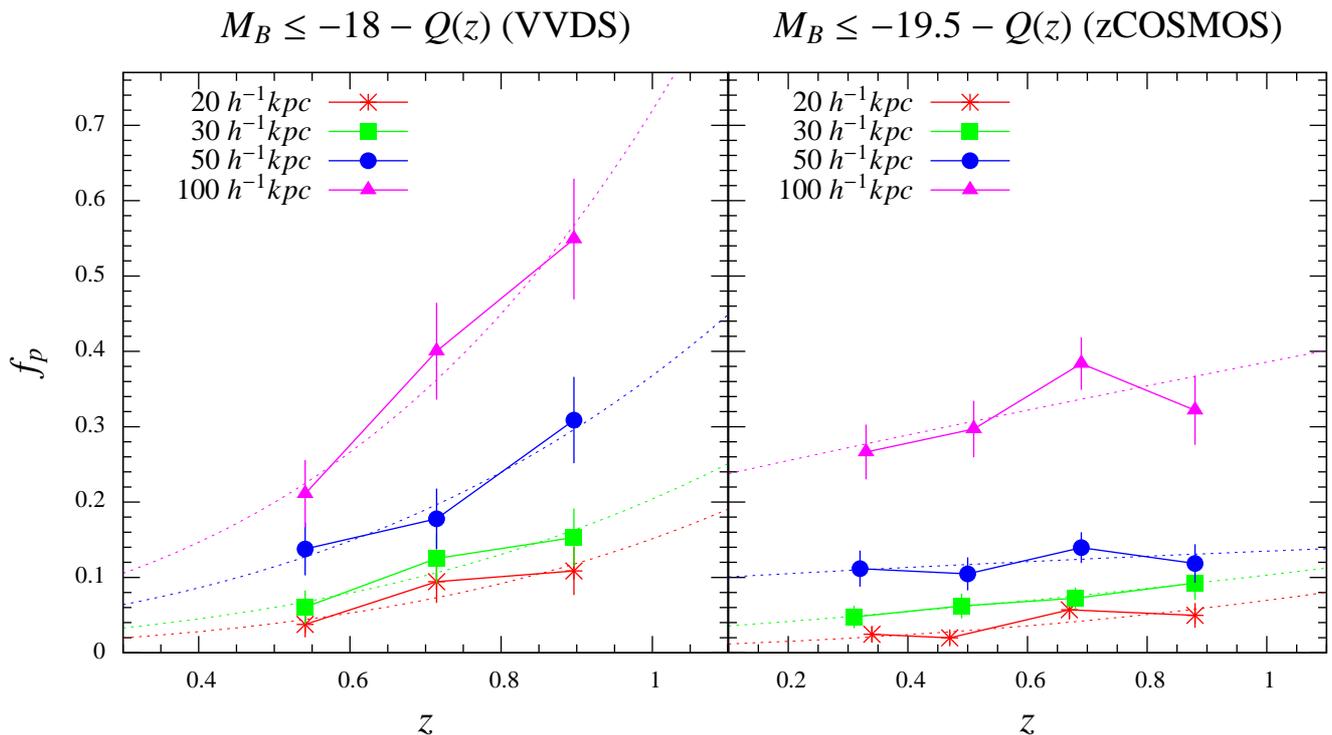}
      \caption{The evolution of the pair fraction with redshift for different sets of projected separations within the VVDS survey (left panel, de Ravel et al., 2009) and zCOSMOS (right panel, this work). Dotted lines are best fits for each separation with the standard parametrization $f_p(z) \propto (1+z)^m$.}
      \label{PF-mag}
 \end{center}
 \end{figure*}

 To investigate the evolution of the pair fraction with redshift we use the standard parametrization $f_p(z) = f_p(z=0) \times (1+z)^m$. A least-squares fit, weighted by the error bars, to our data gives a slope $m$ decreasing from $2.98 \pm 1.62$ with $r_p^{max} = 20\ h^{-1}kpc$ to $0.81 \pm 0.59$ with $r_p^{max} = 100\ h^{-1}kpc$. Pair fractions and corresponding best fits parameters for each separations are showed in Table~\ref{table_PF}. We show in Figure~\ref{PF-mag} the results of best fits for these different separations in the zCOSMOS survey as well as those obtained in \cite{deravel} within the VVDS-Deep sample. The zCOSMOS-\textit{10k} and the VVDS-Deep (\cite{lefevre05a}) are very complementary since they have the same pure I-band selection function but at different depths ($I_{AB}\leq 22.5$ and $\leq 24$ respectively). The VVDS-Deep sample absolute magnitude completeness threshold is $M_B \leq -18$, one magnitude and half fainter than the zCOSMOS-\textit{10k}. Using this two surveys offers the unique opportunity to extend consistently the luminosity/mass ranges of our study. At $z \sim 0.9$, we find a pair fraction for the closest pairs ($r_p^{max} = 20\ h^{-1}\ kpc$) of $5.00 \pm 1.70 \%$ for zCOSMOS galaxies whereas galaxies one magnitude and a half fainter (from the VVDS-Deep) show a pair fraction of $10.86 \pm 3.20 \%$ at the same redshift and for the same separation. Moreover, the general trend shows that the pair fraction evolution of bright galaxies ($m=2.98\pm1.62$) is flatter than those of faint galaxies ($m=4.73\pm2.01$) at any separation criterion up to $z \sim 1$. This indicates that faint galaxies are more likely to be in pairs at $z\sim1$ than bright galaxies. This is not true anymore at $z\sim0.5$ where pair fractions of both population are comparable. Kampczyk et al. (in prep) show that the fractions of spheroid-spheroid pairs, comparable to the so-called \textit{dry}-mergers, in the zCosmos-\textit{10k} sample is $\sim20\%$ at $z\sim0.9$ whereas fraction of early-early type pairs are found to be $\sim3\%$ in the VVDS-Deep. This comparaison supports the fact  that the redshift evolution of the pair fraction up to $z=1$ comes mainly from late type/disc galaxies since their number densities increase at fainter magnitudes.

\begin{table*}[htbp!]
\begin{center}
\caption{Pair fractions for different sets of separations and redshift with $M_B \leq -19.64 - Q(z)$.}
\label{table_PF}
\begin{tabular}{|c | c c c c|}
\hline
& & & & \\
 $z$ bin & $20 h^{-1} kpc$ & $30 h^{-1} kpc$ & $50 h^{-1} kpc$ & $100 h^{-1} kpc$\\  \hline 
 & & & & \\
$0.2 - 0.4$ & $0.025 \pm 0.010$ & $0.047 \pm 0.015$ & $0.112 \pm 0.024$ & $0.267 \pm 0.036$\\  
$0.4 - 0.6$ & $0.020 \pm 0.009$ & $0.062 \pm 0.016$ & $0.105 \pm 0.022$ & $0.297 \pm 0.037$\\  
$0.6 - 0.8$ & $0.057 \pm 0.013$ & $0.072 \pm 0.014$ & $0.140 \pm 0.020$ & $0.383 \pm 0.035$\\  
$0.8 - 1.0$ & $0.050 \pm 0.017$ & $0.093 \pm 0.022$ & $0.119 \pm 0.026$ & $0.322 \pm 0.046$\\
 & & & & \\
\hline\hline

\multicolumn{5}{|c|}{Best fits for standard parametrization $f_p(z) = f_p(0) \times (1+z)^m$}\\  \hline
 & & & & \\
 $f_p(z=0)$ & $0.0088 \pm 0.0072$ & $0.030 \pm 0.002$ & $0.096 \pm 0.003$ & $0.22 \pm 0.06$\\
 $m$ & $2.98 \pm 1.62$ & $1.78 \pm 0.53$ & $0.49 \pm 0.58$ & $0.81 \pm 0.59$\\
 & & & & \\
\hline
\end{tabular}
\end{center}
\end{table*}

\subsection{... in mass selected samples}
\label{pair-fraction-mass}

\begin{figure*}[htbp!]
  \begin{center}
  \input{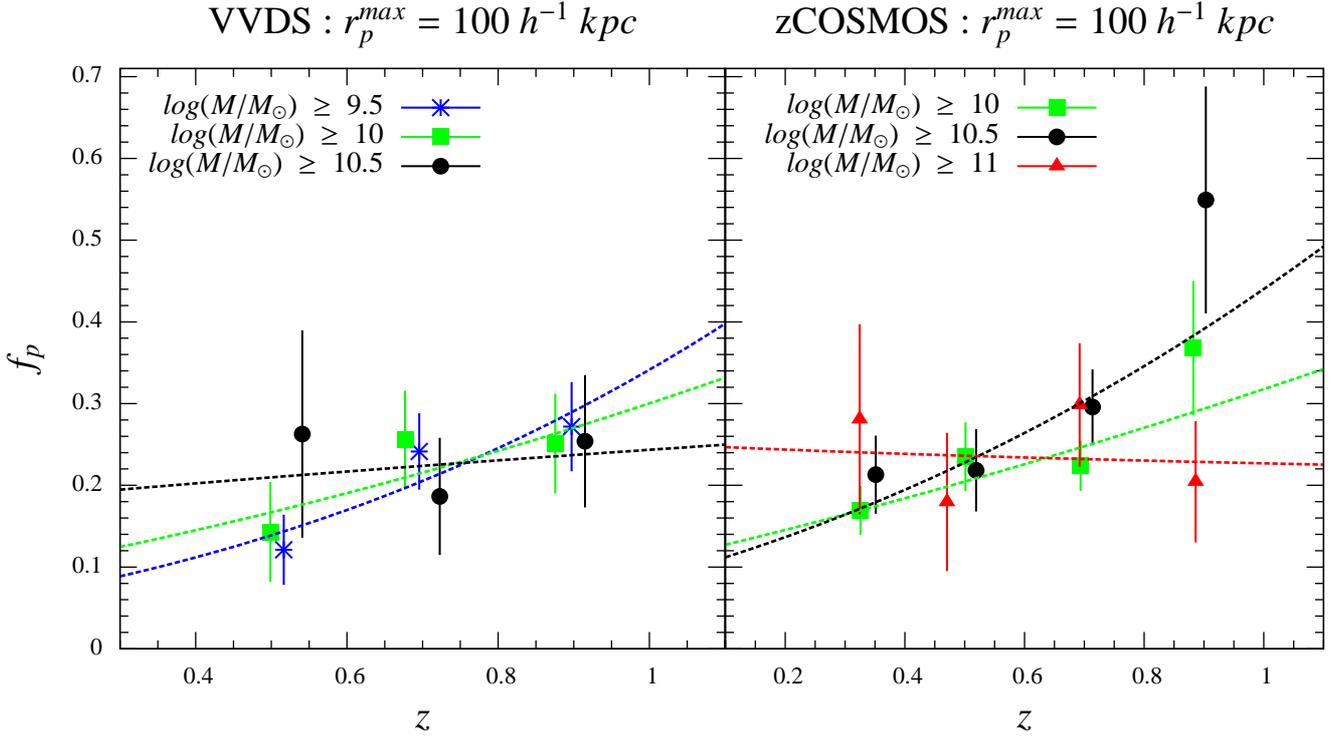}
      \caption{The evolution of the pair fraction with redshift for different mass selected samples within the VVDS survey (\textit{left}, de Ravel et al., 2009) and zCOSMOS (\textit{right}, this work). Dotted lines are best fits for each mass selection with the standard parametrization $f_p(z) \propto (1+z)^m$.}
      \label{PF-mass}
 \end{center}
 \end{figure*}

In this section, we study the influence of the stellar mass on the galaxy pair fraction/merger rate. Since many studies have shown a clear evidence of the influence of luminosity on the merger rate (for instance \cite{P08}; \cite{deravel}; \cite{Lopez09a}; \cite{Lopez09b}; \cite{Bridge10}), we want to understand how this dependence translates in terms of stellar masses. Stellar masses used for this study have been derived using the Bruzual \& Charlot (2003) population synthesis models with an exponentially declining star formation history (SFH) and initial mass function (IMF) of Chabrier et al (2003). More details about the method used to derive stellar masses in the zCOSMOS-\textit{10k} sample can be found in Bolzonella et al. (2009). We are then able to assign a stellar mass to both members of our pairs. In order to select major mergers, we constrain the mass ratio between the two galaxies to be less than 1:4 (corresponding roughly to our absolute magnitude difference in the B-band of 1.5). We then derived for each redshift bin, the limit in stellar mass, $M_{lim}$, above wich the sample is at least 95\% complete for the global population. Depending on the lower limit $M_{min}$ of stellar mass  studied ($log (M_{min}/M_{\odot{}}) = 10,\ 10.5$ or $11$), we derive a correction based on $M_{lim}$ wich takes into account the dependence of the minimum detected stellar mass with redshift and mass-to-light ratio (\cite{pozzetti09}). Following a similar method as described in section \ref{selection_effects}, we derived the pair fractions for different mass selected samples. In order to study the influence of stellar mass on the pair fraction, we had to split our sample therefore diminishing the average  number of pairs available for each case. We therefore show here results for pair fractions estimated from pairs with $r_p^{max} = 100\ h^{-1}\ kpc$ to keep enough statistics. Doing so, 205, 149 and 54 pairs  match the stellar mass criteria $log(M/M_{\odot{}})\ \geq 10,\ 10.5$ and $11$ respectively. Figure~\ref{PF-mass} shows these results for different mass selected samples and compare them with the similar study done in the VVDS-Deep survey (\textit{left} panel). Pair fractions values and best fits for evolution are shown in Table~\ref{table_PF_mass}.

  \begin{table*}[htbp!]
\begin{center}
\caption{Pair fractions for different mass selected samples selected with $r_p^{max} = 100\ h^{-1}\ kpc$.}
\label{table_PF_mass}
\begin{tabular}{|c | c c c|}
\hline
& & & \\
 $z$ bin & $log(M/M_{\odot{}})\ \geq 10$ & $log(M/M_{\odot{}})\ \geq 10.5$ & $log(M/M_{\odot{}})\ \geq 11$\\  \hline 
 & & & \\
$0.2 - 0.4$ & $0.17 \pm 0.03$ & $0.21 \pm 0.05$ & $0.28 \pm 0.12$\\  
$0.4 - 0.6$ & $0.24 \pm 0.04$ & $0.22 \pm 0.05$ & $0.18 \pm 0.08$\\  
$0.6 - 0.8$ & $0.22 \pm 0.03$ & $0.30 \pm 0.05$ & $0.30 \pm 0.08$\\  
$0.8 - 1.0$ & $0.37 \pm 0.08$ & $0.55 \pm 0.14$ & $0.20 \pm 0.07$\\
 & & & \\
\hline\hline

\multicolumn{4}{|c|}{Best fits for standard parametrization $f_p(z) = f_p(0) \times (1+z)^m$}\\  \hline
 & & & \\
 $f_p(z=0)$ & $0.11 \pm 0.04$ & $0.09 \pm 0.04$ & $0.25 \pm 0.15$\\
 $m$ & $1.53 \pm 0.74$ & $2.29 \pm 0.98$ & $-0.14 \pm 1.20$\\
 & & & \\
\hline
\end{tabular}
\end{center}
\end{table*}

Best fits using the standard parametrization are shown for the new results presented here and the VVDS-Deep. One can see a fairly good agreement between the two studies on the overlapping selections ($log(M/M_{\odot{}})\ \geq 10,\ 10.5$) except maybe for the highest redshift bin of $log(M/M_{\odot{}})\ \geq 10.5$ selected galaxies. The zCOSMOS data show a pair fraction of $f = 0.55 \pm 0.14$ at this redshift, higher than the one found in the VVDS-Deep ($f = 0.25 \pm 0.08$) implying a steeper evolution with redshift, but still statistically consistent. Comparing now the two extreme selected sample, one can see a clear difference of trend for low-mass galaxies ($log(M/M_{\odot{}})\ \geq 9.5$) and massive galaxies ($log(M/M_{\odot{}})\ \geq 11$). The pair fraction for low mass galaxies raises from $0.12\pm0.04$ to $0.27\pm0.05$ between $z\sim 0.5$ and $z\sim0.9$ implying an evolution with $m=3.13\pm1.54$ while massive galaxies pair fraction changes from $0.28\pm0.12$ at $z\sim0.3$ to $0.20\pm0.07$ at $z\sim0.9$ implying an evolution with $m=-0.14 \pm 1.2$, consistent with a constant evolution. Therefore we conclude that massive galaxies are less likely to be in pairs at $z\sim1$ and this become less and less true as cosmic time passes and the fraction of elliptical galaxies raises. 

Pozzetti et al. (2009) inferred from the evolution of the galaxy stellar mass function upper limits for the contribution of mergers to the mass accretion process. They found that the stellar mass accretion in intermediate mass range is mainly dominated by accretion due to star formation history. However, after accounting for SFH mass growth they found differences of $20$ to $40 \%$ between the observed stellar mass function and the evolved stellar mass function. Different processes are needed to explain the stellar mass growth of galaxies. Among them, mergers could be a key ingredient. In that sense, these values can be seen as upper limits to the contribution of mergers to mass accretion. In particular, they found a limited evolution of the stellar mass function above $log(M/M_{\odot{}}) = 11$ which suggests that these galaxies formed and assembled their mass before $z=1$. This trend is consistent with our findings that most of the merging activity below $z=1$ appears in low mass regimes.

\section{Galaxy merger rate and stellar mass densities involved}
\label{deriving-merger-rate}
	\subsection{From pair fraction to merger rate}
	\label{merger-rate-estimation}

\begin{figure*}[ht!]
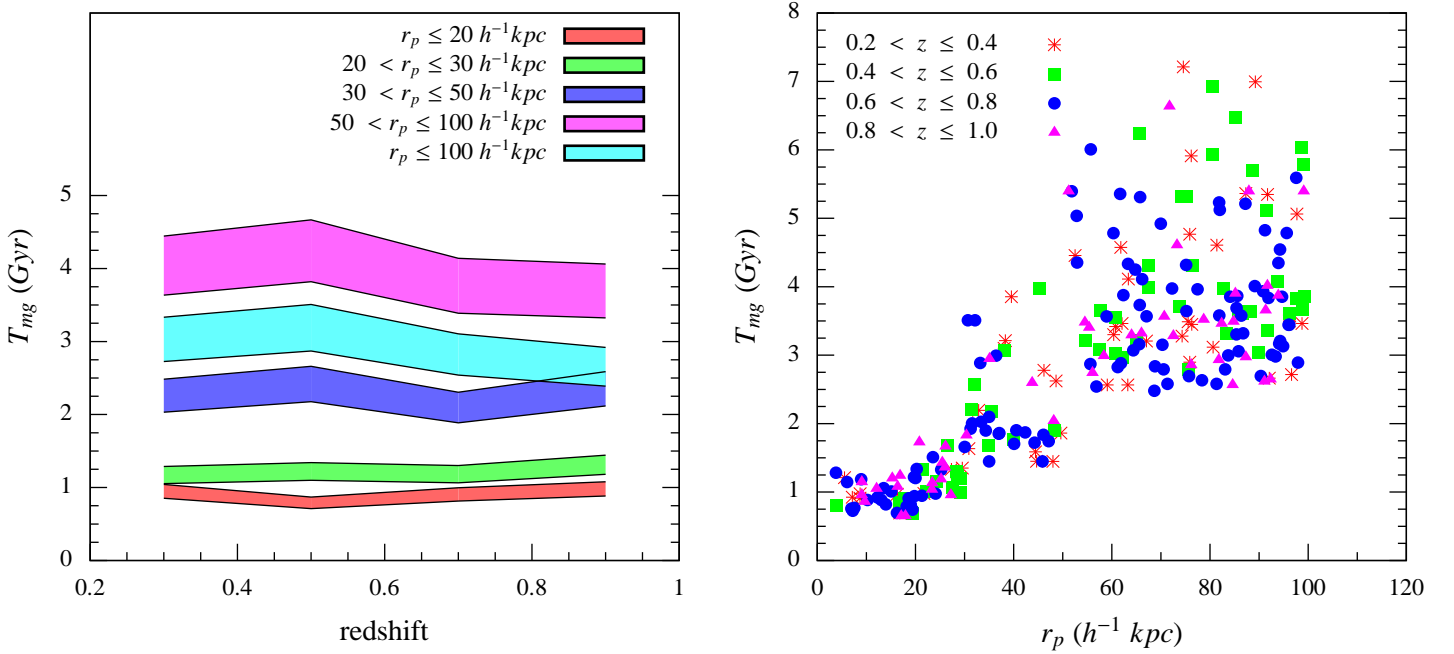

   \begin{center}
     \begin{tabular}{cc}
       \input{Tmg_vs_redshift.tex} &
       \input{Tmg_vs_dr.tex} \\
     \end{tabular}
     \caption{Left Panel : Evolution of merging timescales with redshift for different projected separations $r_p$. Right panel : Merging timescales of galaxy pairs as a function of separation $r_p$ for different redshift bins.}
     \label{time}
   \end{center}
 \end{figure*}

The pair fraction gives an estimate of the number of mergers with respect to the underlying spectroscopic sample. Here, we aim to estimate the merger rate as the number of mergers per unit of time and comoving volume. To this effect we need to estimate mainly two main parameters : $n(z)$, the comoving number density of galaxies in each redshift bin, and $T_{mg,\ k}(r_p^k,z^k)$, the merging timescale of two galaxies with a projected separation of $r_p^k$ at the mean redshift $z^k$. We take results from the Millennium simulations  (\cite{kitz}) to estimate the merging time-scale $T_{mg,\ k}(r_p^k,z^k)$, written as follows: $$T_{mg,\ k}^{-1/2} = T_0(r_p^k)^{-1/2} + f_1(r_p^k) \times z^k + f_2(r_p^k) \times (logM_{*}^k-10),$$ where $M_{*}^k$ is the stellar mass of the primary galaxy. We computed $T_0, f_1$ and $f_2$ for several ranges of projected separations in the case of $\Delta v^{max} = 500\ km/s$, as shown in Figure~\ref{time} (left panel). One can therefore assign to each pair a prediction for its merging timescale as shown in Figure~\ref{time} (right panel). At the largest separations studied here ($r_p^{max} = 100\ h^{-1} kpc$), the mean merging timescale at all redshift is $\sim 3-3.5\ Gyr$ while for the closest pairs this timescale is $\sim 1\ Gyr$. Estimating the merging time-scales is certainly the largest source of uncertainty when trying to translate pair fractions to merger rates. Depending on the method one use, estimates can vary by one order of magnitude (See Section 6.3).

Assuming that the merging time-scale is known, the merger rate is then expressed as : 
\begin{eqnarray}
\centering
N_{mg}(z) = C_{mg} \times f_p(z) \times n(z) \times T_{mg}^{-1}(z)\ , 
\end{eqnarray}
where $T_{mg} (z) =1/N_p \times \sum_k T_{mg,\ k}(r_p^k,z^k)$ is the averaged merging timescale in the redshift bin considered and $C_{mg}$ stands for the merging probability between the two galaxies. This probability takes into account the fact that even when spectroscopic redshift of both members of the pair is known, projection effects can still play a role. $C_{mg}$ has been estimated by Patton \& Attfield (2008) as a function of luminosity using a combination of the SDSS and simulated mocks extracted from the Millennium simulation. They found that in the local Universe, the projection effects contamination tend to increase towards fainter absolute magnitude being  $\sim 50\%$ for bright galaxies and $\sim 70-80\%$ for galaxies below $M_*$, leading to a mean value of $\sim 60\%$.  Lin et al. (2008) have put this constant to $C_{mg}=0.6$ and in order to be consistent for comparaisons, we decided to use the same value. One has to keep in mind that in principle this probability should be a function of projected separation $dr$, luminosity and also environment. Kampczyk et al. (in prep) have studied in details the impact of environment on the projection effects that affect pair studies. At the end of this section, we will estimate the range of probability distributions if one include a projected separation dependence in $C_{mg}$.

Here, we use $C_{mg}=0.6$ which is the value used in de Ravel et al. (2009), and timescales from Kitzbichler \& White (2008). In our analysis, the ratio $C_{mg}/T_{mg}$ depends on the separations of merger candidates (both projected and radial) and on the redshift. A recent study from Lin et al. (2010) have shown that this ratio could also evolved with the local density. However, they found only a weak dependancy of timescales with environment. On the other hand, they found that $C_{mg}$ could evolve from $\sim 0.8$ in low density regions down to $\sim 0.4$ in the higher densities probed by their analysis. They found that both $C_{mg}$ and $T_{mg}$ are only marginally dependant of redshift (up to $z=1.2$). They used a projected separation of $r_p^{max} = 50\ h^{-1} kpc$ to select their merger candidates and found a mean merging timescale of $\sim 1Gyr$. In our analysis, we find that for  $r_p^{max} = 50\ h^{-1} kpc$, the typical merging timescale is $\sim 1.8 Gyr$ which leads to a $C_{mg}/T_{mg}$ ratio of $\sim 0.33$ ($\sim 0.38$ up to z=1), about half the ratio found by Lin et al. ($C_{mg}/T_{mg}$ = 0.7). This comparaison show that depending on the way one treat the merging timescales/probabilities, there is a factor two uncertainty in the pair fraction / merger fraction translation. However, the values used in this studies are consistent with the work done on previous studies on the VVDS and data are therefore directly comparable.

\ \\
The previous definition of the merger rate gives then the number of merger events by unit of comoving volume and time, hereafter we will call it the \textit{volumetric} merger rate. This number gives us an information about the global contribution of mergers to the mass assembly. We can look at the merger rate from a different angle : the \textit{normalized merger rate}. This quantity is the number of merger events by unit of time \textit{per galaxy}. This other approach of the merger rate, normalized to the overall population fitting the selection criterium, enable us to study a potential differential evolution for galaxies of different mass/luminosity. In the following, we will then give the merger rates values expressed from these two point of view, the \textit{volumetric} and the \textit{normalized} merger rates.

\begin{figure*}
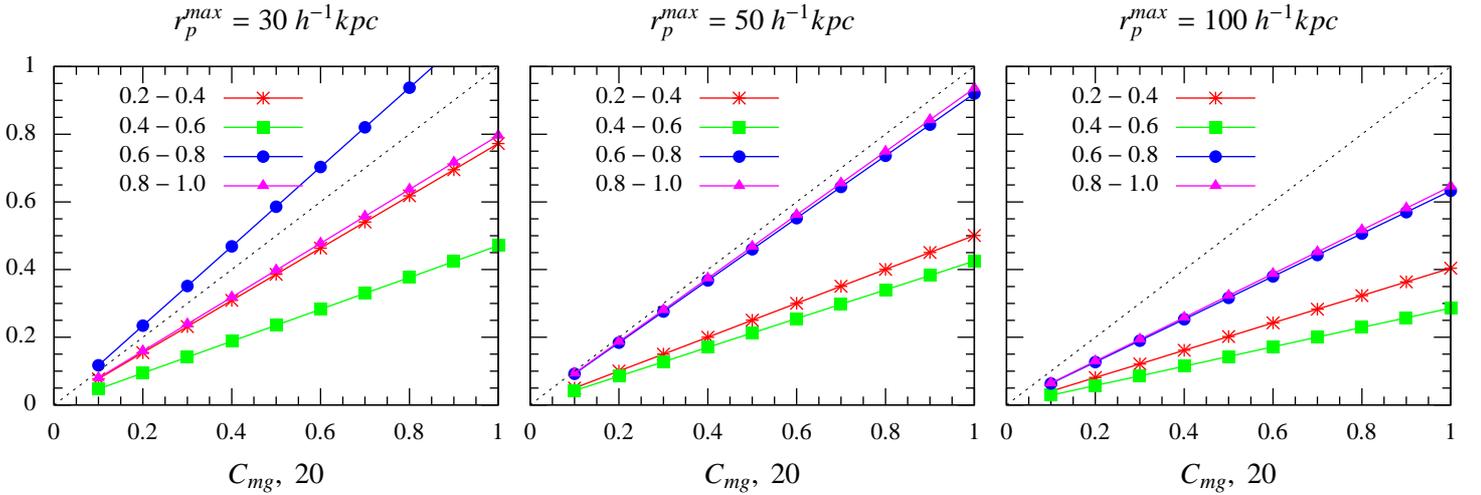

   \begin{center}
     \begin{tabular}{ccc}
       \input{C_mg-30.tex}&\input{C_mg-50.tex}&\input{C_mg-100.tex}\\
    \end{tabular}
     \caption{Probability of merging as a function of $C_{mg,\ 20}$ for different redhift bins. Panels from left to right give the probability to merge for various separations ($30,\ 50$ and $100\ h^{-1} kpc$) as a function of $C_{mg,\ 20}$}. The black dotted line is the same in each panel and is just plotted as an eye-guide.
     \label{prob}
   \end{center}
 \end{figure*}

\ \\
Following de Ravel et al. (2009), we have set $C_{mg}$ constant at $0.6$. 
Obviously, the pair fraction depends on separation but when translating the pair fraction to the merger rate one should get an absolute value. Indeed, different $r_p^{max}$ values should give the same merger rates. In Equation (1) the timescale is independent of $C_{mg}$ and takes into account  the separation effect and since the number density is not linked to the separations,  the fluctuations we observe in the merger rate should mainly come from the fact that $C_{mg}$ is put as a constant. Adding a $r_p$ dependence on $C_{mg}$, one should get the same merger rates for the different separations. A good approximation is to assume the value of $C_{mg} (r_p = 20 h^{-1}\ kpc)$ since most timescales assumptions in simulations have been computed  with this separation. It is then required that $N_{mg,\ 20} = N_{mg,\ \alpha}$ for $\alpha$ going from $30$ to $100 h^{-1}\ kpc$. Therefore, setting $C_{mg} (r_p = 20 h^{-1}\ kpc) = C_{mg,\ 20}$, following Equation (1) the probability for a pair with a given separation $r_p = \alpha$ to merge can be written as : $$ C_{mg,\ \alpha}= C_{mg,\ 20} \times \frac{f^{20} \times T_{mg,\ \alpha}}{f^{\alpha} \times T_{mg,\ 20}},$$
where $f^{\alpha}$ is the pair fraction with the given separation $r_p = \alpha$ leading to a timescale of $T_{mg,\ \alpha}$ . We show in Figure~\ref{prob} the evolution of $C_{mg,\ \alpha}$ with $C_{mg,\ 20}$. 

These estimates give us a range of merging probabilities as a function of $r_p$ and $z$. For instance, if one put $C_{mg,\ 20} = 0.6$, the merging probability for a pair with similar stellar masses is $~0.30-0.70$ at $r_p = 30\ h^{-1}kpc$, $~0.25-0.55$ at $r_p = 50\ h^{-1}kpc$ and $~0.15-0.4$ at $r_p = 100\ h^{-1}kpc$ depending on redshift.
Fluctuations seen in merger rates could therefore be explained by introducing a dependence in $C_{mg}$ with separation and redshift. The \textit{a posteriori} values we found for $C_{mg} (r_p,\ z)$ are reasonnable and follow a consitent trend. This ensure us that, at least qualitatively, the numbers we derived using \textit{Millennium} simulations are consistent.

	\subsection{Influence of luminosity on the merger rate}
	\label{luminosity}

We investigate here the dependency of the evolution of the merger rate on luminosity. 
We give values of \textit{volumetric} merger rates for the luminosity selected sample in Table~\ref{table_Merger}. In order to quantify the evolution with redshift of this merger rate, we use the standard parametrization $N_{mg}(z) = N_{mg} (0) \times (1+z)^{m_{mg}}$. We performed a least-squares fit to our data and best fit parameters are also given in Table~\ref{table_Merger}. Depending on separation, we found that the evolution index $m_{mg}$ evolves from $(1+z)^{2.79 \pm 2.14}$ with $r_p^{max} = 20\ h^{-1}kpc$ to $(1+z)^{0.27 \pm 0.92}$ with $r_p^{max} = 100\ h^{-1}kpc$. Since merger rate should not depend on the projected separation, we can combine the different best fit values found for the different separations and estimate that the mean merger rate evolves as $(2.04 \pm 0.83) (1+z)^{0.97 \pm {0.82}}$ ($\times 10^{-4}$ mergers $h^3\ Mpc^{-3}\ Gyr^{-1}$). \textit{Normalized} merger rates (i.e. the number of mergers per unit of time per galaxy) at different redshifts are also given in Table~\ref{table_Merger_norm}.

\begin{table}[htbp!]
\begin{center}
 \caption{\textit{Volumetric} merger rates for different sets of separations and redshift with $M_B \leq M_B^{lim}$ in units of $\times 10^{-4}$ mergers $h^3\ Mpc^{-3}\ Gyr^{-1}$.}
\label{table_Merger}
\begin{tabular}{|c | c c c c|}
\hline
& & & & \\
 $z$ bin & $20 h^{-1} kpc$ & $30 h^{-1} kpc$ & $50 h^{-1} kpc$ & $100 h^{-1} kpc$\\  \hline 
  & & & & \\
$0.2 - 0.4$ & $1.87 \pm 0.46$ & $2.43 \pm 0.46$ & $3.74 \pm 0.48$ & $4.63 \pm 0.38$\\  
$0.4 - 0.6$ & $1.05 \pm 0.28$ & $2.24 \pm 0.36$ & $2.48 \pm 0.31$ & $3.69 \pm 0.28$\\  
$0.6 - 0.8$ & $3.39 \pm 0.46$ & $2.90 \pm 0.34$ & $3.69 \pm 0.32$ & $5.37 \pm 0.29$\\  
$0.8 - 1.0$ & $2.73 \pm 0.54$ & $3.42 \pm 0.49$ & $2.91 \pm 0.38$ & $4.22 \pm 0.36$\\
 & & & & \\
\hline\hline

\multicolumn{5}{|c|}{Best fits for $N_{mg}(z) = N_{mg}(0) \times (1+z)^{m_{mg}}$ ($\times 10^{-4}$)}\\  \hline
 & & & & \\
 $N_{mg}(0)$ & $0.55 \pm 0.59$ & $1.53 \pm 0.34$ & $3.16 \pm 1.65$ & $3.92 \pm 1.75$\\
 $m_{mg}$ & $2.79 \pm 2.14$ & $1.23 \pm 0.72$ & $-0.03 \pm 1.07$ & $0.27 \pm 0.92$\\
 & & & & \\
\hline
\end{tabular}
\end{center}
\end{table}

\begin{table}[htbp!]
\begin{center}
 \caption{\textit{Normalized} merger rates for different sets of separations and redshift with $M_B \leq M_B^{lim}$ in units of $\times 10^{-2} mergers\ h^3\ Gyr^{-1}\ gal^{-1}$.}
\label{table_Merger_norm}
\begin{tabular}{|c | c c c c|}
\hline
& & & & \\
 $z$ bin & $20 h^{-1} kpc$ & $30 h^{-1} kpc$ & $50 h^{-1} kpc$ & $100 h^{-1} kpc$\\  \hline 
  & & & & \\
$0.2 - 0.4$ & $1.98 \pm 0.48$ & $2.56 \pm 0.48$ & $3.95 \pm 0.51$ & $4.89 \pm 0.40$\\ 
$0.4 - 0.6$ & $1.55 \pm 0.41$ & $3.30 \pm 0.52$ & $3.66 \pm 0.46$ & $5.44 \pm 0.41$\\ 
$0.6 - 0.8$ & $4.43 \pm 0.61$ & $3.79 \pm 0.45$ & $4.82 \pm 0.42$ & $7.01 \pm 0.38$\\ 
$0.8 - 1.0$ & $3.79 \pm 0.75$ & $4.75 \pm 0.69$ & $4.04 \pm 0.52$ & $5.86 \pm 0.50$\\ 
 & & & & \\
\hline
\end{tabular}
\end{center}
\end{table}

\begin{figure*}
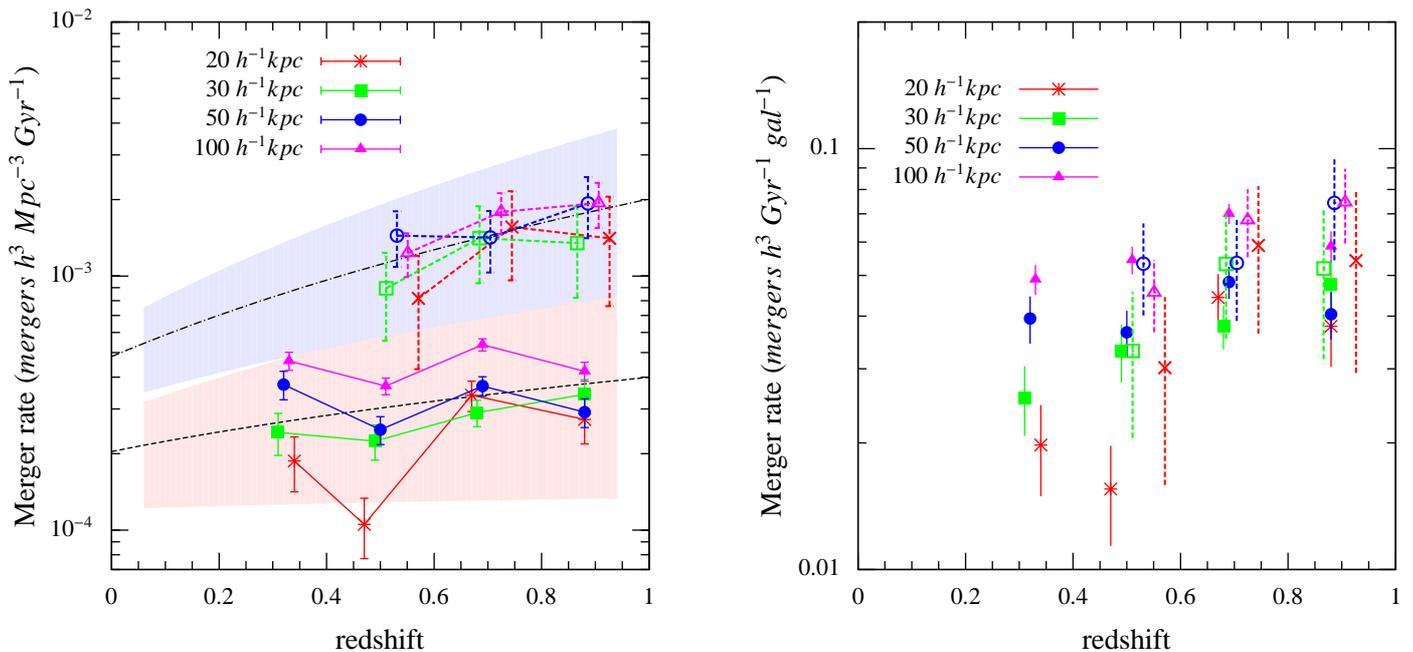

   \begin{center}
     \begin{tabular}{cc}
       \input{merger_rate_comparaison.tex} & \input{merger_rate_comparaison_per_galaxy.tex} \\
    \end{tabular}
     \caption{\textit{Left panel} : Evolution of the volumetric galaxy merger rate with redshift for different projected separations in the zCosmos \textit{10k}-sample (solid lines).We show in dotted lines the merger rates derived from the VVDS-Deep survey ($I_{AB} \leq 24$) one magnitude and half deeper.\textit{Right panel} : Evolution of the merger rate per galaxy in the zCosmos \textit{10k}-sample (solid lines). Dotted lines show again the comparaison with the VVDS-Deep survey.}
     \label{merger-rate}
   \end{center}
 \end{figure*}

We show in Figure~\ref{merger-rate} the evolution of both \textit{volumetric} and \textit{normalized} merger rates for various projected separations in the zCosmos \textit{10k}-sample for galaxies with $M_B\leq -19.64-Q(z)$. We compare the zCOSMOS results to measurements derived from the VVDS-Deep sample (\cite{deravel}).  Within the \textit{10k}-sample ($I_{AB} \leq 22.5$), the mean merger rate of bright galaxies evolves as $(2.04 \pm 0.83) (1+z)^{0.97 \pm {0.82}}$ ($\times 10^{-4}$ mergers $h^3\ Mpc^{-3}\ Gyr^{-1}$). We estimated in the same way the mean merger rate derived within the VVDS-Deep ($I_{AB} \leq 24$) and found that it evolves as $(4.82 \pm 1.62) (1+z)^{2.06 \pm {0.62}}$ ($\times 10^{-4}$ mergers $h^3\ Mpc^{-3}\ Gyr^{-1}$) for faint galaxies. Comparing these results, we find that the evolution of the merger rate is steeper in fainter samples. This confirms and extends to brighter luminosities the trend seen in \cite{deravel}. Integrating these volumetric merger rates over time between $z=0$ and $z=1$ is a way to quantify the contribution of different galaxy populations to the global merger history in the Universe. We find that galaxies brighter than $M_B=-18-Q(z)$ contribute $\sim 3.8$ times more to the global merger history than galaxies brighter than $M_B = -19.6 - Q(z)$ up to $z=1$. 

Figure~\ref{merger-rate} (\textit{right panel}) shows the normalized merger rates per galaxy in the two samples. The first thing one can notice is that the number of mergers per galaxy has decreased by a factor of 2 between $z \sim 0.9$ and $z \sim 0.3$, evolving from about $\sim 0.06$ to $\sim 0.03$ respectively. We don't see significant effects of the luminosity on this evolution. The faint galaxies seem to show a higher \textit{normalized} merger rate but this trend, according to the size of error bars (reflecting poissonnian errors), does not have a high statiscal significance. This means that a given galaxy, regardless of its luminosity, will experiment two times less merging processes between $z=1$ and $z=0.3$.  

	\subsection{Influence of stellar mass on the merger rate}

\begin{figure*}[htbp!]
 \centering
       \resizebox{!}{10.5cm}{\scalebox{1}{\input{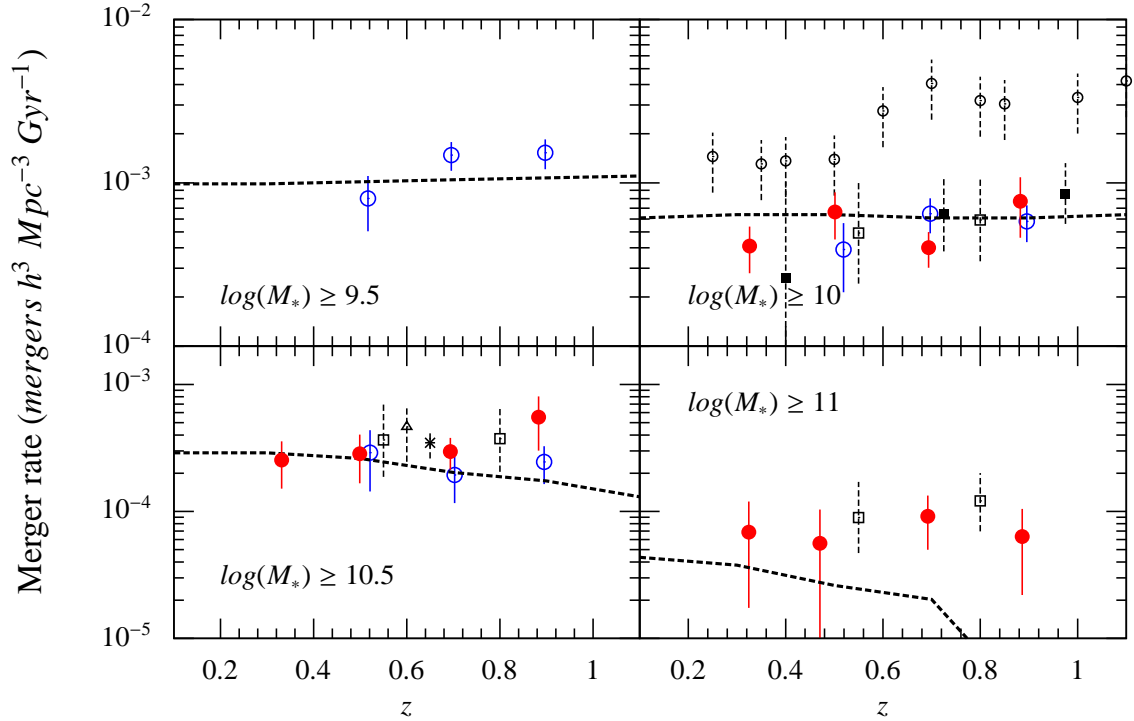}}}
     \caption{Evolution of the volumetric merger rate as a function of redshift for different mass-selected samples from zCosmos (red filled circles) and VVDS (blue empty circles). We also plot several previous studies in black: Conselice et al.,2009 (empty circles), Lopez-Sanjuan et al., 2009a (empty triangle), Lopez-Sanjuan et al., 2009b (filled squares), Bell et al., 2006a (cross), Bundy et al., 2009 (empty squares) and Kitzbichler \& White, 2008 (dotted line)}
     \label{merger-mass}
 \end{figure*}

Figure~\ref{merger-mass} shows the \textit{volumetric} galaxy merger rate as a function of redshift for the different mass-selected samples. For intermediate masses ($log(M/M_{\odot{}}) \geq 10$ and $10.5$) we find good agreements with results from the VVDS. This enables us to compare the lowest ($log(M/M_{\odot{}}) \geq 9.5$) and highest ($log(M/M_{\odot{}}) \geq 11$) mass regimes directly. 

There appears a clear difference of \textit{volumetric} merger rate trends between high and low mass galaxies. We find that low mass galaxies ($log(M/M_{\odot{}}) \geq 9.5$) are more likely to have a major companion than high mass galaxies ($log(M/M_{\odot{}}) \geq 11$) by a factor of order of $\sim 10$ below $z = 1$. For the less massive sample in the VVDS ($log(M/M_{\odot{}}) \geq 9.5$), we find a merger rate evolving as $N_{mg} (0) \times (1+z)^{m_{mg}}$ with $m_{mg} = 2.38 \pm 1.57$ and $N_{mg} (z=0) = (3.56 \pm 3.17) \times 10^{-4}\ $ mergers$\ h^3 Mpc^{-3} Gyr^{-1}$, while for the massive sample in zCOSMOS-\textit{10k} sample, we find that the merger rate evolves with $m_{mg} = 0.73 \pm 1.80$ and $N_{mg} (z=0) = (0.48 \pm 0.44) \times 10^{-4}\ $ mergers$\ h^3 Mpc^{-3} Gyr^{-1}$. 

Therefore, major mergers below $z = 1$ are found to be more likely for low mass galaxies, merging history of high mass galaxies being roughly low and constant for massive galaxies during the last eight billion years of galaxies evolution. We find an average value of $8 \times 10^{-5}\ mergers\ h^3 Mpc^{-3} Gyr^{-1}$ for our massive sample in really good agreement with the results found by Chou et al. (2010) on the flat evolution of massive dry mergers up to $z=0.7$.

Figure~\ref{merger-mass} also shows a comparison between our estimates of volumetric merger rates and several previous studies which use different methods to infer the galaxy merger rate. 

Kitzbichler \& White (2008) made N-body simulations of DM evolution (Millennium) in combinaison with a semi-analytical model for galaxy formation to explore the relationship between the evolution of galaxy merger rate and observed close pairs. They find that the galaxy merger rate decreases with stellar mass. We find good agreements with their data, except maybe for the most massive galaxies sample, where we find a moderate ( more or less flat ) evolution with redshift while they find a strong decrease.

Bundy et al. (2009) used deep Infra Red observations on the northern GOODS field combined with public surveys in GOODS-S to investigate the dependence on stellar mass of the merger rate up to $z \sim 1.2$. At least up to $z =1$, the highest redshift we can reach here, merger rates are in good agreement.

Lopez-Sanjuan et al. (2009a) measured the fraction of galaxies undergoing disk-disk major mergers at $0.35 \leq z \leq 0.85$ by studying the asymmetry index A of galaxy images. They have corrected with care the bias due to varying spatial resolution and image depth with redshift and developed a method to account for incertainties in asymmetry indices and photometric redshift. Morphological and spectroscopical studies of close pairs are hard to compare since, by definition, the two technics trace mergers at different stages. In the litterature, morphological studies tend to find higher pair fractions than those derived with spectroscopic samples. Nevertheless, we find good agreement with Lopez-Sanjuan et al. (2009a) which can be seen here as a lower limit to the global merger rate, since they studied disk-disk mergers. We also compare our results with Lopez-Sanjuan et al. (2009b). In this paper, they measured the merger rate of $log(M)\geq 10$ galaxies in GOODS-S using also assymetries. Again a nice agreement is found. 

In another morphological based study, Conselice et al. (2009) used deep HST imaging from the Extended Groth Strip and the COSMOS field to examine structural properties (CAS) of galaxy pairs and derived the merger rate evolution for galaxies with $log(M/M_{\odot{}}) \geq 10$ up to $z = 1.2$. Their results are found to be in disagreement with ours by a factor of 2-3 at all redshifts considered here whereas pair fractions seem consistent. Lotz et al. (2008) showed that the time-scale for a $20\ kpc$ pair to merge is $0.2\pm0.1\ Gyr$, significantly lower than what we used in this analysis ($\approx 1 Gyr$). This might be the main explanation for the apparent disprecancy we find in merger rates. Conselice et al. (2009) also investigated the systematic difference between morphological-based and pair count estimates of merger rates. They compared pair fractions derived with pair count (\cite{kartaltepe07}) and morphological criteria and found that this apparent disprecancy (by a factor varying from 1.5 to 3.5 with Kartaltepe et al. (2007) results) comes from the different time-scales probed by the two methods. However, although they used the same approach, our agreement is much better with  \cite{Lopez09b}, raising up an issue between morphologically based analysis themselves. Lopez et al. (2009b) claim that this discrepancy with Conselice et al. (2009) is due to the fact that they don't take into account the observationnal errors contained in the asymmetry parameter $A$ (see Section 6 of their paper). This effect could lead to an over-estimation by a factor $\sim 3$ of the final merging rate consistent with what the comparaison shows.

\begin{figure*}[htbp!]
 \centering
       \resizebox{!}{10.5cm}{\scalebox{1}{\input{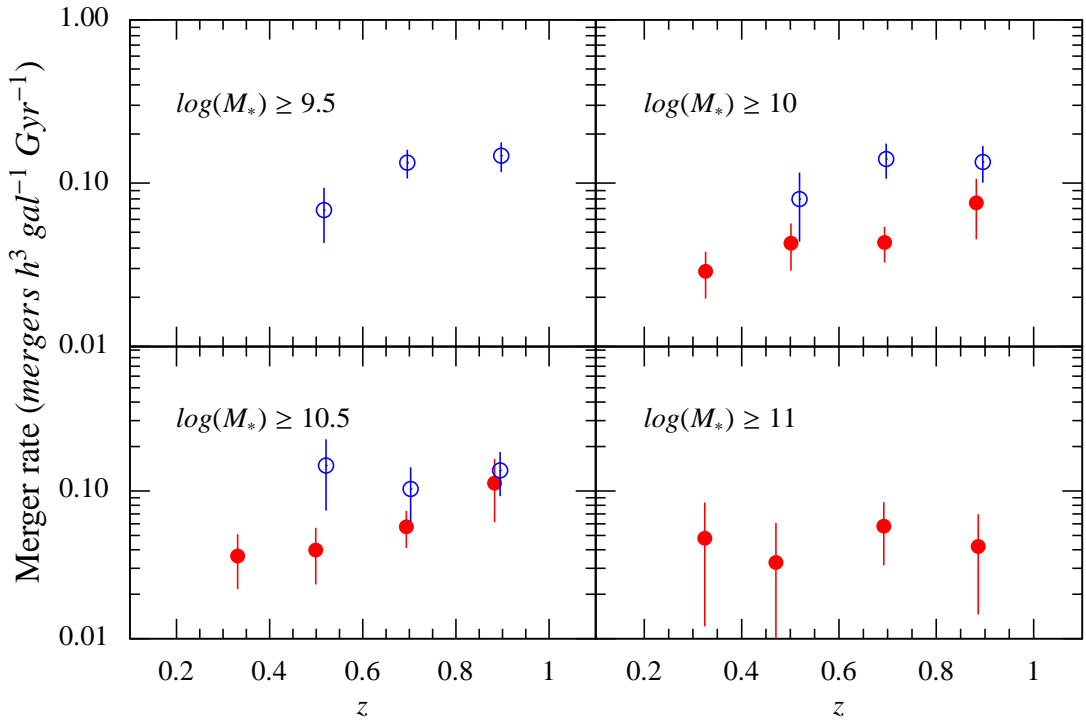}}}
     \caption{Evolution of the merger rate per galaxy as a function of redshift for different mass-selected samples from zCosmos (red filled circles) and VVDS (blue empty circles).}
     \label{merger-mass-per-galaxy}
 \end{figure*}

Figure~\ref{merger-mass-per-galaxy} shows the evolution of the \textit{normalized} merger rate for different mass selected samples in the zCOSMOS and the VVDS surveys. As for the luminosity selected samples, one can see a weakening of the trend when one normalizes to the global population. However this effect is less strong for mass selected samples since massive galaxies ($log(M/M_{\odot{}}) \geq 11$) show between two and three times less merging activity (per galaxy) than less massive galaxies ($log(M/M_{\odot{}}) \geq 9.5$) at $z\sim1$. Indeed, Kitzbichler \& White merging time-scales are found to enhance the merger rate stellar mass dependence while their dependence with luminosity is less strong (Bundy et al., 2009). This could explain the different trends we see here between stellar mass and luminosity selected samples. 
Anyway, this result remains statistically significant and shows that low mass galaxies ($log(M/M_{\odot{}}) \geq 9.5$) are more likely to experience a major merger at high redshift. Moreover, at any redshift below $z=1$, these galaxies show more merging activity than the most massive galaxies in our sample ($log(M/M_{\odot{}}) \geq 11$).

\subsection{Stellar mass density involved in merger processes}
\label{stellar-mass-density}
The \textit{volumetric} merger rate estimates the number of galaxy mergers by unit of time and volume. Assuming that one know the total stellar mass involved in the merging process, this can be translated into the fraction of the stellar mass which is involved in a merger process. For each pair $k$, containing galaxies $i$ and $j$, we are able to define the stellar mass $M_{\displaystyle merger}^{\displaystyle k}\ =\ M_i +M_j$ involved in the merger process. The stellar mass density involved in the merger is then written as : $$\rho_{\displaystyle M}^{\displaystyle merger} (z)=N_{\displaystyle mg}(z) \times T_{\displaystyle bin}(z) \times \frac{\displaystyle 1}{\displaystyle N_{p} (z)} \times \sum_{\displaystyle k=0}^{\displaystyle N_{p}(z)} M_{\displaystyle merger}^{\displaystyle k}(z) ,$$ where $T_{bin}(z)$ is the elapsed time correponding to the redshift bin concerned, $N_p$ the number of pairs.
We estimate the fraction $f_{\displaystyle \rho}$ of the stellar mass density involved in a merging process relative to the underlying population as : $$ f_{\displaystyle \rho}=\frac{\displaystyle \rho_{\displaystyle M}^{\displaystyle merger} (z)}{\rho_{\displaystyle M}(z)} ,$$ where $\rho_{\displaystyle M}(z)$ is the mean stellar mass density of the global population for each redshift bin.

Following Patton et al. (2000), we estimate the fraction of stellar mass that have undergone a major merger since $z \sim 1$. We performed this calculations for mass selected galaxies and found that $12.2 \pm 3.9$, $13.1 \pm 5.6$ and $9.7 \pm 5.2\%$ of stellar mass with $log(M/M_{\odot{}}) \geq 10$, $10.5$ and $11$ has been involved in a merger process since $z=1$. This shows the higher merging activity trend of less massive galaxies below $z=1$. Given the uncertainties on the merging time-scales and the stellar masses themselves, we definitely can't rule out a consistent merging history between those galaxy populations at this point. However,  these numbers can be compared to the number of $\sim 20\%$ for galaxies with $log(M/M_{\odot{}}) \geq 9.5$ found in de Ravel et al. (2009), showing that the major merging activity for today massive galaxies ($log(M/M_{\odot{}}) \geq 11$) since $z = 1$ has two times less important than lower mass galaxies ($log(M/M_{\odot{}}) \geq 9.5$) suggesting an earlier epoch for massive galaxies mass assembly through merging process. Out of the $20$ to $40\%$ of residual evolution found by \cite{pozzetti09} after accounting for star formation history, between $10$ and $15\%$ are due to major mergers. In other words, between $z\sim1$ and $z\sim 0$, our studies suggest that stellar mass accretion of galaxies is due to \textit{in situ} star formation activity at the level of $\sim 70\%$ and major mergers at the level of $\sim 10-15\%$. This leaves $15-20\%$ of evolution that could come from other physical processes like minor mergers or prolongated star formation histories (second bursts). Lopez-Sanjuan et al. (2010b) have shown that the relative contribution of the mass growth by merging is $\sim 75\%\ /\ 25\%$ due to major / minor mergers. In our case this would translate into a minor merger induced mass accretion of $\sim 2.5 - 4\%$ since $z\sim1$.

\section{Impact of environment on the global merger rate}
\label{environment}
\begin{figure}[htbp!]
  \begin{center}
        \resizebox{!}{6.5cm}{\scalebox{1}{\input{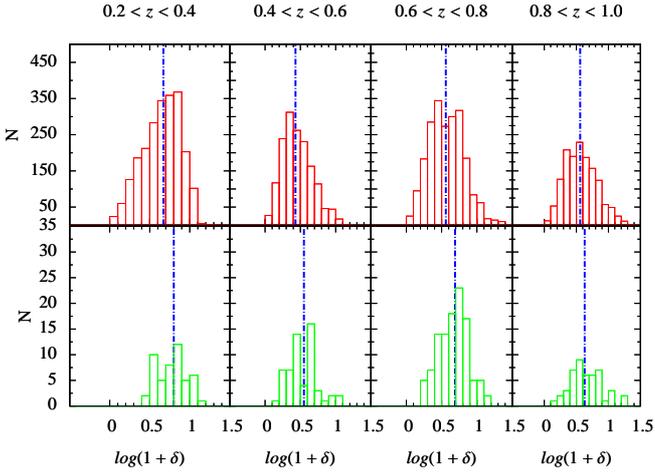}}}\\
      \caption{Over-density distribution of the underlying sample (top panels) and $r_p^{max}=100 h^{-1}kpc$ selected pairs (bottom panel) for several redshift bins. We used fixed aperture ($3 Mpc$) mass weighted densities estimations. The dashed line is the median of the distribution in each case. For all redshift bins, the distribution median is slightly higher for paired galaxies.}
      \label{type-densities}
 \end{center}
 \end{figure}

The galaxy overdensity field in the zCOSMOS area has been studied in details in Kovac et al. (2010). Using this density information, we are able to investigate in which type of environment mergers are preferentially taking place as cosmic time evolves. Many density tracers have been derived to estimate the density, and different tracers of galaxy populations have been used. Here we used the flux limited sample ($I_{AB} \leq 22.5$) which enable us to use our entire sample in order to keep enough statistics. The density computations have been performed either on a fixed scale ( for instance $3\ Mpc/h$) or on variable scales corresponding to the projected distance of the $N^{th}$ nearest neighbour. The mean separation between a galaxy and its $5^{th}$ nearest neighbour has been shown to be always less than $3\ Mpc/h$. This approach traces the highest densities of the field, with the caveat that the scale varies with redshift (from $\sim\ 1\ Mpc/h$ at $z \sim\ 0.2$ to $\sim\ 3\ Mpc/h$ at $z \sim\ 1$). Since we are interested in the evolution of the merger rate with redshift, we decided to use densities computed using a fixed aperture of $3\ Mpc/h$. 
Densities have been computed either using a simple count of galaxies, or using a weight on the luminosity or the stellar mass of galaxies. It is expected that some galaxy properties, and particularly masses, are better tracers of the underlying matter density field, therefore we decided to use the densities computed using a mass weighted scheme. We studied the influence of the weighting scheme on our results and found that the observed trends remain regardless of the weighting scheme used. Therefore we decided to use densities estimated with the flux limited sample, on a fixed scale of $3\ Mpc/h$ and where the density is weighted by the stellar masses of galaxies. In the general case, the density at a given position $r = r(\alpha,\delta,z)$ is given by (\cite{kovac10}): $$\rho(r) =\sum_{i} \frac{m_i\ W(|r-r_i|;R)}{\phi(r_i)}\, $$ where $W$ is a spatial window function, $m_i$ is the weight based on the stellar mass of galaxies, $\phi(r_i)$ a correction function for various observationnal issues and $R$ is the smoothing length. In the following, we will refer to the density (density contrast) $\delta$ defined as : $$\delta (r) = \frac{\rho(r) - \rho_m(z)}{\rho_m(z)}\ ,$$
where $\phi_m(z)$ is the mean density at a given redshift.

\ \\
Figure~\ref{type-densities} shows the density distribution of the underlying sample and the pair samples (selected with $r_p^{max} = 100 h^{-1}\ kpc$ and $M_B \leq -19.64-Q(z)$) within several redshift bins. Galaxy pairs are found to be systematically in denser regions at all redshifts. This is of course expected but to quantify this trend, we studied the evolution of the fraction of pairs with respect to the underlying sample as a function the density $1+\delta$.

\begin{figure}[htbp!]
  \begin{center}
       \resizebox{!}{7.5cm}{\scalebox{1}{\input{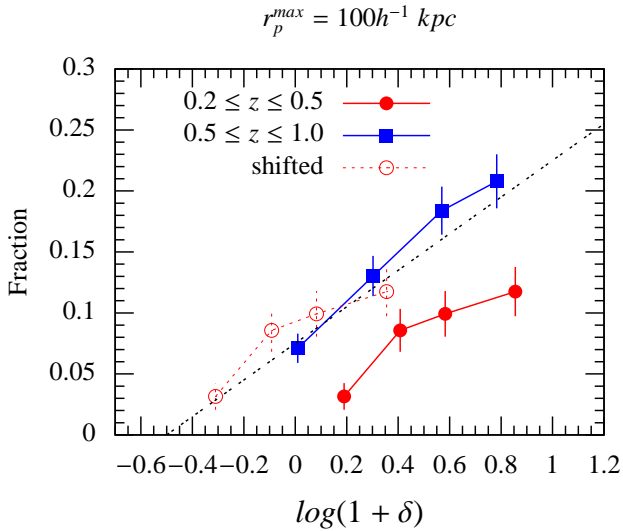}}}\\
      \caption{The fraction of pairs with respect to the underlying sample is plotted versus the over-density for pairs with $r_p^{max} = 100h^{-1}\ kpc$. We show in dotted circles the shift applied to the low redshift sample to correct for the evolution of the mean density of the structures (see text). The dotted line is the best fit we get while fitting the high-redshift and corrected low redhift bins data.}
      \label{PF-densities}
 \end{center}
 \end{figure}

Figure~\ref{PF-densities} shows the pair fraction as a function of the overdensity $\delta$ for $0.2 \leq z \leq 0.5$ and $0.5 \leq z \leq 1$ and for different projected separations $r_p^{max} = 100h^{-1}\ kpc$. 
 
Abscisse values are defined using quartiles of densities distributions. We find that the fraction of pairs is higher in denser environments for both high and low redshift regimes. In order to evaluate the evolution with redshift of this relation, we have corrected the densities from the evolution in the mean density $\delta_{m}$ evolving as $\propto (1+z)^3$, translating to a change in the mean density of $\Delta (log(1+\delta_m)) \sim 0.45$ between the two redshift ranges considered, as plotted in Figure~\ref{PF-densities}. Comparing the 'mean density corrected' relations at low redshift  and high redshift values, we infer that the fraction of pairs depends mainly on the local density with little or no evolution with redshift, consistent with recent results within the DEEP2 survey by Lin et al. (2010). A trend also found in samples with less statistics with $r_p^{max} = 30,\ 50\ h^{-1}\ kpc$. The dependence of the number of pairs to the local density, over the full redshift range considered and up to $z\sim1$ is compatible with a power law evolving like $(1+\delta)^{\sim 0.15}$, $(1+\delta)^{\sim 0.05}$ and  $(1+\delta)^{\sim 0.03}$ for separations of $100$, $50$ and $30\ h^{-1}kpc$, respectively. This means that for an environment 10 times denser, one finds $3.0$, $2.6$ and $2.1$ times more pairs for separations of $r_p^{max}=100$, $50$ and $30\ h^{-1}kpc$ respectively. Finally, quantitatively speaking, the pair fraction is about $2-3$ times higher in environment one order of magnitude denser.

Kampczyk et al. (in prep) have investigated the possible contamination due to objects that fit our pair criteria but will not lead to a merger process, which corresponds to the introduction of a environment dependence in $C_{mg}$. This effect is expected to be stronger in over-dense regions since the gravitationnal bound between objects is weaker. They find that for the highest overdensity quartile this correction can decrease the pair fraction by $~20\%$ for pairs with $dr^{max} = 100\ h^{-1}kpc$ and that it is considerably smaller for lower separations of $dr^{max} \leq 50\ h^{-1}kpc$ (up to $\sim 12\%$) and $dr^{max} \leq 30\ h^{-1}kpc$ (less than $4\%$). For all the other over-density quartile, this effect is almost negligible (see also Lin et al. 2010 for a theoritical perspective).

\begin{figure}[htbp!]
 \begin{center}
    \input{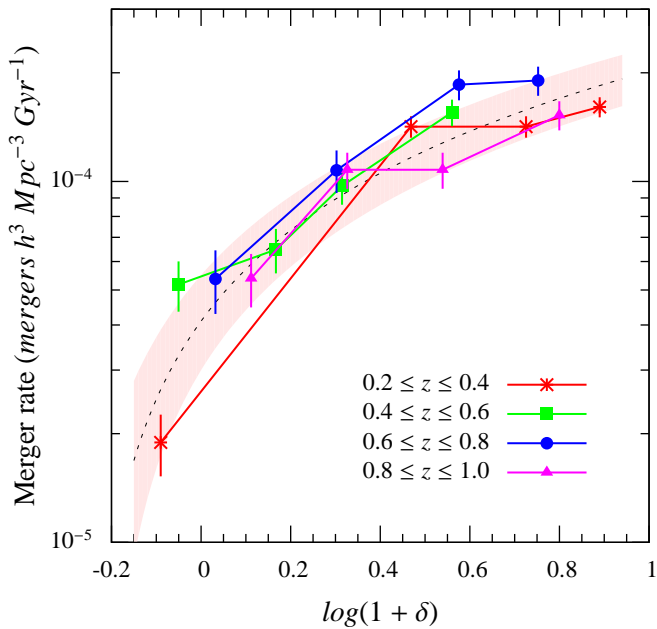}
   \caption{Merger rate as a function of environment for different redshift bins. The dotted indicate the best for all the data and the shaded area the $1-\sigma$ error on the best fit.}
      \label{merger-rate_density}
 \end{center}
  \end{figure}

In order to estimate the contribution of  different density quartiles to the global merger rate we show in Figure~\ref{merger-rate_density}  the merger rate as a function of environment for different redshifts bins using $\leq 100\ h^{-1}\ kpc$ pairs. We believe this estimate is legit because our merger rate estimation takes into account the fact that $\leq 100\ h^{-1}\ kpc$ pairs have higher merging time-scales on average and in theory we should be able to get a merger rate regardless of the separation assumed to define pairs. As expected, most of the global merger rate comes from the high density regions. However, the impact of local density on the merging time-scale is not taken into account here but, as mentionned before, this effect has been estimated to be up to $\sim 20\%$ for the very dense regions. This will lead to a decrease of our merger rate estimate from $\sim 2$ to $\sim 1.6  \times 10^{-4}\ mergers\ h^3 Mpc^{-3} Gyr^{-1}$ in the $0.8 \leq log(1+\delta) \leq 1.0$ regions. 

We find the evolution of the merger rate with environment to be weakly dependant on the redshift up to $z=1$. We performed a least square fit on data to parametrize this evolution. We found that the galaxy merger rate evolves as $N_{mg} (\delta) = a \times log(1+\delta) +b$ with $a = (1.61 \pm 0.13) \times 10^{-4}$ and $b = (0.41 \pm 0.05) \times 10^{-4}$. This means that the number density of merging processes in high density regions is increased by a factor $\sim 8-10$. 

\section{Discussion and conclusions}
\label{conclusion}
We have studied the evolution of the merger rate with redshift in the zCosmos-\textit{10k} bright survey for galaxies brighter than $M_B \leq -19.64 - Q(z)$. We have identified a sample of 39 (263) pairs with each member of the pair having a secure spectroscopic redshift with separations up to ($20 (100)\ h^{-1}kpc,\ 500\ km\ s^{-1}$). With the avaibility of the local galaxy density field measured for this sample, we have been able to investigate the influence of environment on the pair fraction and the merger rate. 
The pair fraction increases from $2.5 \pm 1.0 \%$ at $z \sim 0.3$ to about $5.0 \pm 1.7 \%$ at $z \sim 0.9$ for galaxies selected with $I_{AB} \leq 22.5$ and brighter than $M_B = -19.6 - Q(z)$. When comparing to galaxies also I-band selected in the VVDS-Deep survey, but 1.5 magnitude fainter, with a pair fraction of about $10\%$ at $z \sim 0.9$. (\cite{deravel}), we find that the pair fraction is dependent on luminosity, with bright galaxies showing a lower pair fraction, confirming the de Ravel et al. (2009) result. 
We find that the pair fraction of bright galaxies evolves with redshift as $(1+z)^m$ with $m = 2.98 \pm 1.62$ for separations of $20\ h^{-1}\ kpc$ decreasing to $m=0.81 \pm 0.59$ for separations of $100\ h^{-1}\ kpc$. Using merging timescales derived from the Millennium simulation (\cite{kitz}), this translates to a mean merger rate evolving as $(2.04 \pm 0.83)\ \times\ (1+z)^{0.97 \pm  0.82}\ (\times 10^{-4}\ mergers\ h^3 Mpc^{-3} Gyr^{-1})$ for galaxies brighter than $M_B \leq -19.64 - Q(z)$. The mean merger rate between $z=0$ and $z=1$ is about $3 \times 10^{-4}\ mergers\ h^3 Mpc^{-3} Gyr^{-1})$. This is $\sim 5$ times lower than what has been found for galaxies 1.5 magnitude fainter (\cite{deravel}). 
We therefore confirm that the pair fraction and the \textit{volumetric} merger rate are dependant on luminosity, with fainter galaxies involved in more mergers than brighter galaxies. This results is consistent with what is found in the analysis of the Millenium simulation (\cite{kitz}; \cite{P08}). 
However, when normalized to the global underlying population, the number of mergers per galaxy of both bright and faint population are comparable while less massive galaxies still show 2-3 times more merging activity than massive galaxies at $z\sim1$. This indicates that the density of mergers depends on the depth of the observations, being higher for fainter observations, while the number of mergers a galaxy will experience does not depends strongly on its intrinsic luminosity but rather on its stellar mass. This result indicate that the merging time-scales derived by Kitzbichler \& White (2008) are more sensitive to the stellar mass. This is confirmed when comparing these time-scales to those estimated from Patton \& Atfield (2008) where the stellar mass dependence of time-scales is shown to be more moderate.  

We found a clear difference of merging histories between low and high mass galaxies, \textit{volumetric} merging rates evolving as $N_{mg} (0) \times (1+z)^{m_{mg}}$ with $m_{mg} = 2.38 \pm 1.57$ and $N_{mg} (z=0) = (3.56 \pm 3.17) \times 10^{-4}\ $ mergers$\ h^3 Mpc^{-3} Gyr^{-1}$ for galaxies with ($log (M/M_{\odot{}}) \geq 9.5$), while for the massive sample in zCOSMOS-\textit{10k} sample, we find that the merger rate evolves with $m_{mg} = 0.73 \pm 1.80$ and $N_{mg} (z=0) = (0.48 \pm 0.44) \times 10^{-4}\ $ mergers$\ h^3 Mpc^{-3} Gyr^{-1}$. This leads to a fraction of $\sim 20\%$ ($\sim 9.7\%$) of the today stellar mass with $log (M/M_{\odot{}}) \geq 9.5$ (respectively $\geq 11$) that have been accreted through a major merger since $z =1$.

We have measured the pair fraction as a function of the local environment and we find that galaxy pairs are more likely to be found in high density regions with a probability increasing by a factor of 2-3 when increasing the local density by a factor of 10. The pair fraction is globaly higher at higher redshift at all densities, following the evolution of the mean density of the Universe with cosmic time. As expected, the pair fraction increases faster in high density regions when the pair separation is increased. We have derived the evolution of the merger rate as a function of environment and find that the merger rate is increasing by one order of magnitude when the local density increases by the same amount, parametrized with $N_{mg} (\delta) = a \times log(1+\delta) +b$ with $a = (1.61 \pm 0.13) \times 10^{-4}$ and $b = (0.41 \pm 0.05) \times 10^{-4}$. We find that the values of this parametrization are only marginally different as redshift increases. Recent local measurements of the evolution of the fraction of mergers as a function of environment have shown that the peak of mergers distribution resides in intermediate density regions (\cite{darg09}). These results seem to be qualititavely in agreement with our conclusions on higher redshifts. From these measurements, we therefore conclude that higher environments produce higher merger rates with only a weak dependence with redshift up to $z=1$. This indicates that the environment has a clear and constant effect on the merger rate throughout cosmic time. Galaxies in higher density regions are subject to 10 times more merging events when the density is increasing by a factor of 10 whatever the epoch between $z=1$ and $z=0$. This behaviour is naively expected as in higher density regions the mean separation between galaxies decreases. The link between merger rate and local density is also supported by simulations (\cite{perez}). 
Galaxies in high density regions will therefore accrete mass significantly faster than in low density regions. This is expected to significantly contribute to several properties like the observed differences in the stellar mass function and luminosity function of galaxies in high and low density environments (\cite{bolzo09}; \cite{zucca09}).
This study therefore brings new evidence that the environment plays a key role in shapping up galaxies along cosmic time.

\begin{acknowledgements}
zCOSMOS. This work has been supported in part by the grant ASI/COFIS/WP3110 I/026/07/0.
\end{acknowledgements}

\end{document}